\documentclass[twocolumn]{aastex62}

\newcommand{\fracbrac}[2]{\left(\frac{#1}{#2}\right)}
\newcommand{\dd}[2]{\frac{d#1}{d#2}}
\newcommand{\pd}[2]{\frac{\partial #1}{\partial #2}}

\newcommand{\secref}[1]{Section \ref{#1}}
\newcommand{\figref}[1]{Figure \ref{#1}}

\newcommand{\github}{\href{https://github.com/shadden/ResonantPlanetPairsRVModeling}{github.com/shadden/ResonantPlanetPairsRVModeling}}
\usepackage{natbib,graphicx,amsmath}
\usepackage{multirow}
\usepackage{floatrow}
\newfloatcommand{capbtabbox}{table}[][\FBwidth]
\usepackage{blindtext}
\accepted{16 July 2020}
\submitjournal{AJ}
\shorttitle{RVs of Resonant Planets}
\shortauthors{Hadden \& Payne}
\begin{document}
\title{Modeling Radial Velocity Data of Resonant Planets to Infer Migration Histories}
\author[0000-0002-1032-0783]{Sam~Hadden}
\affiliation{Harvard-Smithsonian Center for Astrophysics, 60 Garden St., MS 51, Cambridge, MA 02138, USA}
\correspondingauthor{Sam~Hadden}
\email{samuel.hadden@cfa.harvard.edu}
\author[0000-0001-5133-6303]{Matthew~J.~Payne} 
\affiliation{Harvard-Smithsonian Center for Astrophysics, 60 Garden St., MS 51, Cambridge, MA 02138, USA}
%%%%%%%%%%%%%%%%%%%%%%%%%%%%%%%%%%%%%%%%%
%%%%%%%%%%%%%%% abstract %%%%%%%%%%%%%%%%
%%%%%%%%%%%%%%%%%%%%%%%%%%%%%%%%%%%%%%%%%
\begin{abstract}
A number of giant-planet pairs with period ratios $\lesssim 2$ discovered by the radial velocity (RV) method  may reside in mean motion resonances. 
Convergent orbital migration and resonant capture at the time of formation would naturally explain the present-day resonant orbital configurations of these systems.
Planets that experience smooth migration and eccentricity-damping forces due to a protoplanetary disk should not only be captured into mean motion resonances but also end up in a specific dynamical configuration within the resonance, sometimes referred to as apsidal corotation resonance (ACR).
Here we develop a method for testing the hypothesis that a planet pair resides in an ACR by directly fitting RV data. 
The ACR hypothesis strongly restricts the number of free parameters describing the RV signal and we compare fits using this highly restricted model to fits using a more conventional two-planet RV model by using nested sampling simulations.
We apply our method to HD 45364 and HD 33844, two systems hosting  giant-planet pairs in 3:2 and 5:3 resonances, respectively.
{The observations of both systems are consistent with ACR configurations, which are formally preferred based on the Bayes factors computed from nested sampling simulations. 
We use the results of our ACR model fits to constrain the possible migration histories of these systems.}
% These planets are not traditionally expected to form in such close proximity and therefore are thought to have experienced post-formation migration. 
\end{abstract}
%%%%%%%%%%%%%%%%%%%%%%%%%%%%%%%%%%%%%
%%%%%%%%%%%%%% intro %%%%%%%%%%%%%%%%
%%%%%%%%%%%%%%%%%%%%%%%%%%%%%%%%%%%%%
\section{Introduction} 
\label{sec:intro}
    The potential for planets to experience large-scale migration due to interactions with their protoplanetary disk has been understood for decades \citep{GT80,LinPapaloizou1986}.
    However, due to the theoretical uncertainties and numerical difficulties associated with simulating realistic protoplanetary disks, rates and even the directions of planet migration remain uncertain and planet--disk interactions have remained an active area of research \citep[][]{Kley2012,Baruteau2014}. 
    % one robust prediction of 
    It is well known that convergent orbital migration generally causes migrating planets to capture into mean motion resonances \citep[MMRs; e.g.,][]{Goldreich1965} and, despite the theoretical uncertainties, disk-induced migration is expected to produce resonant orbital configurations in at least some planetary systems during their formation.
    Exoplanetary systems hosting planet pairs in MMRs are generally interpreted as the by-products of  orbital migration since such resonant configurations are unlikely to arise by chance.
    Systems with resonant orbital configurations are therefore ideal targets for better understanding the role of planet--disk interactions in shaping the architectures of planetary systems.   

    The discovery of two giant planets in a 2:1 MMR around GJ 876 \citep{Marcy2001} provided the first observed example of an exoplanetary system in resonance around a main-sequence star.\footnote{The pulsar planets orbiting PSR 1257+12 can claim the simultaneous distinctions of being the first exoplanetary system discovered \citep{Wolszczan1992} and the first (near-)resonant exoplanetary system discovered \citep{Malhotra1992}.}
    Subsequent studies showed that the configuration of this resonant system could be reproduced
    by convergent migration in a disk \citep{Snellgrove2001,LeePeale2002}.
    There have since been numerous examples of radial-velocity (RV)-discovered planetary systems hosting closely spaced pairs of massive planets with period ratios $\lesssim2$ that are in or suspected to be in MMR \citep[e.g.][]{Mayor2004,Lee2006,Tinney2006,Correia2009,Niedzielski2009,Johnson2011_VI,Johnson2011_VII,Wright2011,Robertson2012,Wittenmyer2014,Wittenmyer2016,Giguere2015,Luque2019,Trifonov2019}.
    
    Most previous studies have concluded that specific systems occupy resonant configurations by conducting postfitting analyses. This has usually involved fitting RV data using a Markov chain Monte Carlo (MCMC) approach, then initializing $N$-body integrations with initial conditions drawn from their MCMC posterior samples, and finally analyzing the behavior of resonant angles in these $N$-body integrations.
    The goal of this paper is to develop a different approach to analyzing the evidence for MMRs among RV planets: we ask whether the observed RV signals of a given planet pair can be adequately explained by a model that presumes the system resides in a resonant configuration reached by smooth migration and eccentricity damping. 
    More precisely, we assess the evidence for whether a given system resides in a particular dynamical MMR configuration referred to as an `apsidal corotation resonance' or ACR (the libration of the relative position of the periastra is referred to as apsidal corotation).
    ACRs are the natural outcome of resonant capture under the influence of  migration and eccentricity-damping forces that planets are expected to experience in a disk. 
    The benefit of this approach is that it allows us to use RV observations as a direct test of a precisely formulated, falsifiable hypothesis rather than simply interpreting possibly resonant configurations as circumstantial evidence of past migration. 
    Planet pairs that are consistent with an ACR configuration support the hypothesis that the system was shaped by planet--disk interactions.
    Moreover, the specifics of the dynamical state of such systems can be used to constrain the migration and eccentricity damping caused by those interactions.
    Conversely, resonant planet pairs exhibiting RVs inconsistent with an ACR configuration would demand an explanation for their current configuration beyond simple capture by smooth migration such as forcing by turbulent fluctuations of the protoplanetary disk \citep{Adams2008} or gravitational interactions with additional planets.

   \citet{Baluev2008} and \citet{Baluev2014} have previously adopted the approach of fitting RV signals while constraining resonant planets to reside in an ACR configuration. 
    We expand upon their methods by incorporating an ACR model into  Bayesian inference and model comparison frameworks using MCMC and nested sampling simulations.
    We also use fits with these models to constrain planet--disk interactions and infer the migration histories of the systems.
    \citet{Delisle2015} previously attempted to constrain planet--disk  interactions in systems hosting resonant giant planets, though they used previously reported orbital solutions rather than re-fitting RVs assuming an ACR configuration. 
    {\citet{Silburt2017} studied the pair of resonant planets in the HD 155358 system using a Bayesian model comparison framework to infer migration parameters capable of reproducing the observed system, though many of the details of their method differ from ours.}

    Our paper is structured as follows: \secref{sec:capture} reviews the dynamics of resonant capture and ACRs. 
    In \secref{sec:rv}, we present our 
    method for Bayesian model comparison and show that the two systems we consider are consistent with an ACR configuration.
    We discuss our results in \secref{sec:discussion} including inferences about the possible migration histories of the systems and the prospects for better determining the resonant state of the planets with future observations. 
    Finally, \secref{sec:conclusions} summarizes our conclusions.

%%%%%%%%%%%%%%%%%%%%%%%%%%%%%%%%%%%%%%%%%%%%%%%%%%%%%%%%%%%%%%%%%%%%%%
%%%%%%%%%%%%%%%%%%%%%% Resonance Capture %%%%%%%%%%%%%%%%%%%%%%%%%%%%%
%%%%%%%%%%%%%%%%%%%%%%%%%%%%%%%%%%%%%%%%%%%%%%%%%%%%%%%%%%%%%%%%%%%%%%
\section{The Dynamics of Resonance Capture} 
\label{sec:capture}
\begin{figure*}
    \centering
    \includegraphics[width=\textwidth]{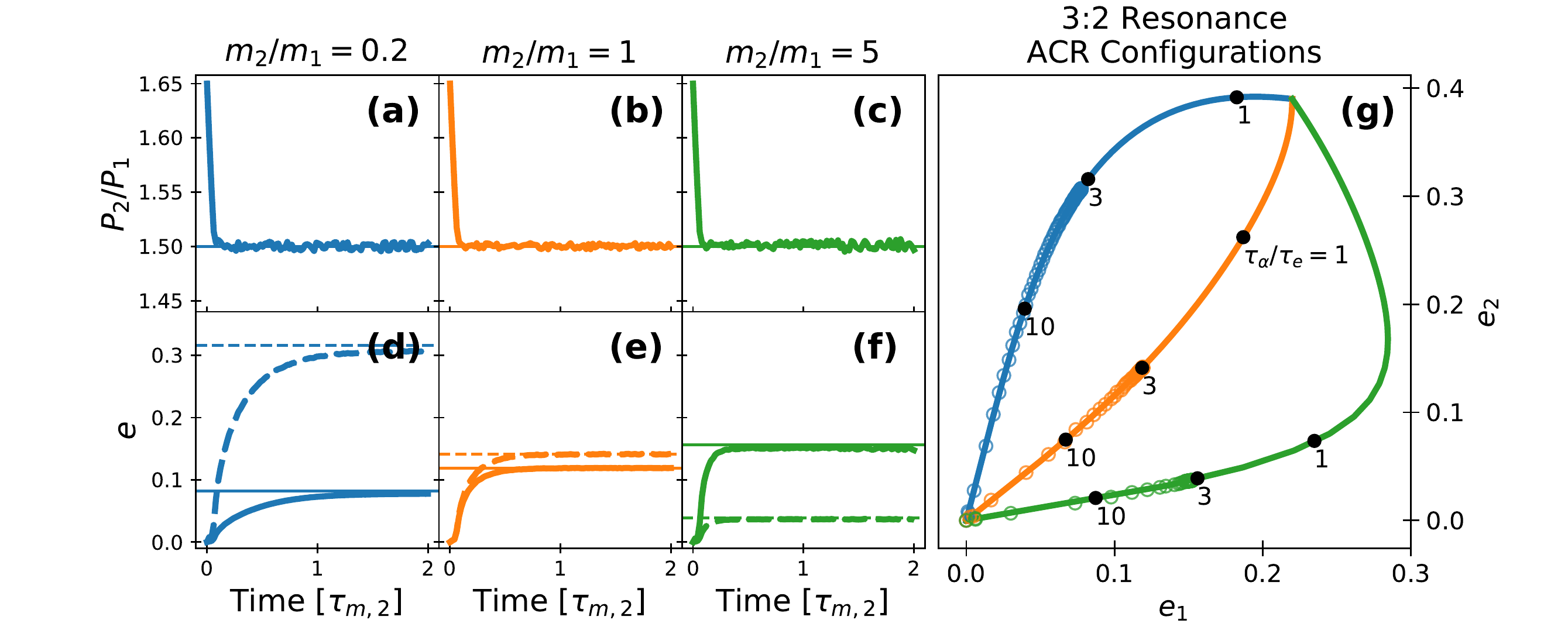}
    \caption{
    Panels (a)--(f) show $N$-body migration simulations for pairs of planets with $m_1 + m_2= 5\times10^{-4}M_*$ and different mass ratios. 
    The simulations are initialized with coplanar planets on initially circular orbits with  $P_2 = 1.1\times\frac{3}{2}P_1$ and random initial orbital phases.
    Dissipative forces are included according to Equation \eqref{eq:capture:dissipation_forces} with $\tau_{e,1}=\frac{m_2}{m_1}\tau_{e,2}=3\times 10^{3}P_1$,  $\tau_{m,2} = 3\tau_{e,1}$, $\tau_{m,1}=\infty$, and $p=0$.
    Panels (a)--(c) show the period ratios and panels (d)--(f) show the eccentricities of the planets. 
    In panels (d)--(f), the eccentricity of the outer planet is shown as a dashed curve while the eccentricity of the inner planet is shown as a solid curve. Equilibrium values of the eccentricities predicted with our semianalytic equations of motion are plotted as thin horizontal lines.
    In panel (g), colored curves show inner and outer planet eccentricities along families of ACR configurations of the 3:2 MMR for the different planet-mass ratios used in the simulations shown in panels (a)--(f). 
    Black points indicate dissipative equilibria for different ratios of $\tau_\alpha/\tau_{e,1}$ with $\tau_{e,2} = (m_1/m_2)\tau_{e,1}$.
    Results of $N$-body migration simulation with $\tau_\alpha/\tau_{e,1} =3$  are shown as circles.
    See text for additional details.
    }
    \label{fig:capture:acr_tracks}
    \end{figure*}

    The capture of migrating planets into MMR is a well-studied problem  \citep[e.g.][]{Goldreich1965,Yoder1973,Henrard1983,Peale1986,MD1999ssd}.
    Migrating planets will be captured into an MMR when the net effect of their migration drives them toward one another, provided the migration is sufficiently slow and the planets' eccentricities are small at the time they encounter the resonance.
    Upon capture into resonance, the planets' period ratio remains fixed at the resonant  period ratio while their eccentricities grow. 
    The planets' eccentricities will continue to grow until they escape the resonance or, if the planets are also subject to eccentricity-damping, the resonance forcing is counterbalanced by the eccentricity damping forces. 
    Figure \ref{fig:capture:acr_tracks} shows illustrative $N$-body simulations of this process for a selection of different planet-mass ratios.
    The $N$-body integrations were carried out with the \texttt{rebound} \citep{rebound} code using the \texttt{IAS15} integrator \citep{ias15} with the precision parameter set to its default value, $\epsilon_b=10^{-9}$. 
    Eccentricity-damping and migration forces were included using the \texttt{modify\_orbits\_direct} 
    routine of the \texttt{reboundx} package \citep{reboundx}.
    All subsequent $N$-body integrations presented in this paper are carried out in the same manner unless explicitly stated otherwise.

    In the examples shown in Figure \ref{fig:capture:acr_tracks}, the dissipative forces quickly drive the system toward an equilibrium configuration where the planets' period ratio and eccentricities remain fixed while the planet pair continues to migrate together in lockstep.
    The particular equilibrium configuration reached by a migrating planet pair can be predicted by examining the equations of motion governing the resonant dynamics of the system. 
    We give a brief overview of these equations here; a complete description is given in Appendix \ref{app:Equations}. 

    To study the dynamics of a $j$:$j-k$ resonance, we adopt a Hamiltonian formulation of the equations of motion in terms of canonical coordinates $\sigma_i = (1+s) \lambda_2 - s\lambda_1 - \varpi_i$, where $s=(j-k)/k$, and their conjugate momenta $ I_i\sim  e_i^2$.\footnote{
    Strictly speaking, our equations describe the dynamical evolution of the planets' {mean} elements. 
    These differ from the planets' osculating elements by a small amount, of order $(m_1+m_2)/M_*$.
    In Appendix \ref{app:Equations}, we distinguish between mean and osculating elements and derive the transformation between the two sets of orbital elements. 
    For simplicity, we will ignore the distinction in the main text.
}
    We use a numerical averaging procedure described in Appendix \ref{app:Equations} to compute the resonant Hamiltonian and its derivatives, and consequently, the validity of our study is not limited to small eccentricities (unlike studies that employ truncated series expansions to approximate the Hamiltonian).
    For compactness, we  write the dynamical variables of our Hamiltonian system as a vector ${\bf z}\equiv (\sigma_1,\sigma_2, I_1, I_2)$. 
    The governing Hamiltonian, ${\bar H}({\bf z};{\cal D})$, depends on the planet--star mass ratios, $m_i/M_*$,  as well as a parameter, ${\cal D}$, the normalized angular momentum deficit (AMD), which is approximately given by
\begin{multline}
    {\cal D} \approx 
    \beta_1\sqrt{\alpha}\frac{e_1^2}{2} + \beta_2\frac{e_2^2}{2}-\frac{k\beta_2\beta_1\sqrt{\alpha}}{3 \left(j\beta_1\sqrt{\alpha}+(j-k)\beta_2\right)}\Delta
\end{multline}
    where $\beta_i=m_i/(m_1+m_2)$, $\alpha = a_{1}/a_{2}$, and $\Delta = \frac{j-k}{j}\frac{P_2}{P_1}-1$.
    For planets deep in resonance  such that $\Delta \approx 0$, ${\cal D}$ can be interpreted as a mass-weighted measure of the planets' eccentricities.  
    $\cal D$ is conserved by the resonant dynamics in the absence of dissipation.
    
    We also consider dissipative forces with a simple parameterized model that imposes migration and eccentricity damping as
    \begin{eqnarray}
    \label{eq:capture:dissipation_forces}
        \frac{d\ln e_i}{dt} &=& -\frac{1}{\tau_{e,i}}\nonumber\\
        \frac{d\ln a_i}{dt} &=& -\left(\frac{1}{\tau_{m,i}}+\frac{2pe_i^2}{\tau_{e,i}}\right)\equiv-{\frac{1}{\tau_{a,i}(e_i)}},
\end{eqnarray}
\noindent{}where $p$ is a parameter that couples eccentricity damping to the semi-major axis damping, with $p=1$ corresponding to eccentricity damping at constant angular momentum \citep[e..g,][]{Goldreich2014,Deck2015}.

The equations of motion governing the dynamics of the resonance are
\begin{eqnarray}
\label{eq:capture:equations_of_motion}
    \dot {\bf z} &=& \Omega \cdot \nabla_{{\bf z}} {\bar H}({\bf z};{\cal D}) + {\bf f}_\text{dis}({\bf z},{\cal D})\nonumber\\
    \dot {\cal D} &=& f_{\text{dis},{\cal D}}({\bf z},{\cal D})
\end{eqnarray}
where 
\begin{eqnarray*}
\Omega &=& \begin{pmatrix}
 0 & \mathbb{I}_2\\
 -\mathbb{I}_2 & 0 
\end{pmatrix}\\
{\bf f}_{\text{dis}}({\bf z},{\cal D})&=& -\sum_{i=1}^2\left(\frac{1}{\tau_{e,i}}\pd{{\bf z}}{\ln e_i}+\frac{1}{\tau_{a,i}}\pd{{\bf z}}{\ln a_i}\right)\nonumber\\
f_{\text{dis},{\cal D}}({\bf z},{\cal D})&=& -\sum_{i=1}^2\left(\frac{1}{\tau_{e,i}}\pd{{\cal D}}{\ln e_i}+\frac{1}{\tau_{a,i}}\pd{{\cal D}}{\ln a_i}\right)\nonumber~.
\end{eqnarray*}
    The equilibrium configuration for a given set of planet masses and dissipation timescales can be determined by finding stationary solutions to Equation \eqref{eq:capture:equations_of_motion}.
    Provided the dissipation timescales are long compared to the characteristic timescales of the resonant dynamics, the equilibrium configuration will be close to a fixed point of the dissipation-free resonant dynamics, 
    i.e., a point satisfying $\nabla_{\bf z} {\bar H}({\bf z};{\cal D})= 0$.
    % Only elliptic equilibria
    These fixed-point configurations of the resonant dynamics have been the subject of numerous studies \cite[e.g.,][]{LeePeale2002,Beauge2003,Haghighipour2003,Michtchenko2006,Antoniadou2014} and are frequently referred to as apsidal corotation resonances (ACRs).
    {As equilibrium points of a Hamiltonian system with two degrees of freedom, 
    ACRs can be either elliptic equilibria, where the Jacobian matrix of the equations of motion has two purely imaginary eigenvalue pairs, or (partially) hyperbolic equilibria where at least one eigenvalue pair has a nonzero real parts. However, because hyperbolic equilibria are unstable to arbitrarily small perturbations, only elliptic ACR equilibria represent plausible dynamical configurations for real planetary systems and hereafter references to ACRs should be implicitly understood as referring only to these stable equilibria.
    }
    These configurations are characterized by zero libration amplitudes of the pair's resonant angles and antialignment of planets' apsidal lines.\footnote{
    Some ACR configurations, such as those of  $N$:1 MMRs, can exhibit aligned apsidal lines or even asymmetric configurations.
    {Resonant capture into the 3:2 and 5:3 MMRs considered in this paper leads to antialigned ACR configurations, so we restrict our considerations to these configurations.}
    }
    For fixed planet masses, ACR configurations of a given MMR form a one-parameter family of orbital configurations that can be parameterized by $\cal D$, the normalized AMD \citep{Delisle2015}.

    \figref{fig:capture:acr_tracks} shows the equilibrium eccentricities for the ACR configuration of the 3:2 MMR for three different planet masses in panel (g).
    Upon encountering the resonance, the migrating planet pairs' period ratios become locked at the resonant value and their AMD increases, with their eccentricities following along the family of ACR configurations as shown in \figref{fig:capture:acr_tracks}.  
    The planets continue to follow these ACR tracks until they reach an equilibrium set by a balance between the strengths of the forcing causing migration and eccentricity-damping so that $\dot{\cal D}=0$.
    Locations of these equilibria for different ratios of the  migration timescale,\footnote{Resonant dynamics depend only on a planet pair's semi-major axis ratio and not the absolute scale of the planets' semi-major axes, which merely determines the timescale of the evolution. Therefore, the ACR equilibrium reached by a planet pair depends on the planets' individual migration timescales $\tau_{m,i}$ only through the combination $\tau_{\alpha}$.} $\tau_{\alpha} = (\tau_{m,2}^{-1} - \tau_{m,1}^{-1})^{-1}$, to the eccentricity-damping timescales, $\tau_{e,i}$, evaluated at different planet mass ratios, are illustrated in \figref{fig:capture:acr_tracks} by the series of black points indicating equilibrium configurations for different ratios of $\tau_\alpha/\tau_{e,1}$ with $\tau_{e,2} = (m_1/m_2)\tau_{e,1}$.

    Depending on the planets' mass ratio and the details of the eccentricity damping mechanisms, the equilibrium configuration reached by the planets may not be stable: upon reaching the equilibrium, the resonant libration amplitude can actually grow \citep{MeyerWisdom2008,Goldreich2014,Deck2015,Delisle2015,Xu2017,Xu2018}. 
    This growth continues until the the planets escape the resonance or the system reaches a limit cycle and their resonant librations saturate at a finite value.
    The stability of a particular equilibrium configuration can be determined by computing the Jacobian matrix,
    \begin{equation}
    \label{eq:capture:jac}
        J = \begin{pmatrix}
        \Omega \cdot \nabla_{\bf z}^2  H + \nabla_{\bf z} {\bf f}_\text{dis} 
        &
         \pd{}{\cal D}\left[\Omega \cdot \nabla_{\bf z} {\bar H} + {\bf f}_\text{dis}\right]^{T}
        \\
        \nabla_{\bf z} f_{{\cal D},\text{dis}}
         & 
         \pd{f_{{\cal D},\text{dis}}}{\cal D}
        \end{pmatrix}~,
    \end{equation}
    of Equations \eqref{eq:capture:equations_of_motion}.
    The stability of the equilibrium is then determined by the eigenvalues of $J$.
    At an equilibrium near an {elliptic fixed point of the dissipation-free system}, $J$ will have two eigenvalues of the form $\pm i\omega_j +\rho_j$ with $|\rho_j|\sim 1/\tau_\alpha$. 
    If either $\rho_j>0$, then the equilibrium is an unstable spiral and any small displacement from equilibrium will be amplified on a timescale $1/\rho_j$.
    In this paper, we will determine the stability of equilibrium configurations 
    by numerical eigenvalue determination after evaluating the Jacobian in Equation \eqref{eq:capture:jac} by means of automatic differentiation.  
    We refer the reader to previous studies for approximate analytic equilibrium stability criteria  \citep{Goldreich2014,Deck2015,Delisle2015,Xu2017,Xu2018}.
    %

%%%%%%%%%%%%%%%%%%%%%%%%%%%%%%%%%%%%%%%%%%%%%%%%%%%
%%%%%%%%%%%%%%%%% Fitting Methods %%%%%%%%%%%%%%%%%
%%%%%%%%%%%%%%%%%%%%%%%%%%%%%%%%%%%%%%%%%%%%%%%%%%%
\section{Radial Velocity Fitting}
\label{sec:rv}
    Resonant planet pairs that owe their dynamical configuration to smooth convergent migration in their protoplanetary disk should reside in or very near an ACR configuration. 
    In Section \ref{sec:rv:methods} we describe an RV-fitting method for testing the hypothesis that a given set of RVs are produced by a system residing in an ACR configuration.
    We then apply this method in Section \ref{sec:rv:data} to two RV systems hosting pairs of giant planets, 
    HD 45364  (§ \ref{sec:rv:data:HD45364}) and HD 33844 (§ \ref{sec:rv:data:HD33844}).
    We show that these systems' RVs are consistent with ACR configurations and that ACR configurations are formally preferred in both cases.  
\subsection{Methods}
\label{sec:rv:methods}
    Our first goal is to develop a method to evaluate whether a planetary system's RV observations can be explained by a pair of planets in an ACR configuration. 
    In order to do so, we adopt a Bayesian model comparison framework \citep[e.g.,][]{gregory2005bayesian}. 
    Given a set of observational data $D$ and a model ${\cal M}_i$ for how that data depend on a set of model parameters $\theta$, Bayes theorem states
    \begin{equation*}
        p(\theta|D,{\cal M}_i) = \frac{p(D|\theta,{\cal M}_i)p(\theta|{\cal M}_i)}{p(D|{\cal M}_i)}~.
    \end{equation*}
    In standard Bayesian parlance, $p(\theta|D,{\cal M}_i)$ is referred to as the posterior, $p(D|\theta,{\cal M}_i)$ as the likelihood, $p(\theta|{\cal M}_i)$ as the prior, and $p(D|{\cal M})$ as the evidence. Parameter inference methods such as MCMC compute or approximate the posterior distribution, $p(\theta|D,{\cal M}_i)$, which represents one's beliefs about the underlying system parameters, conditioned on the observation.
    In model comparison problems, one is instead interested in determining evidences, $p(D|{\cal M}_i)$, for two or more competing models.
    Two models ${\cal M}_i$ and ${\cal M}_j$ can be compared by computing their Bayes factor, ${p(D|{\cal M}_i)
    }/   {p(D|{\cal M}_j)}$, the ratio of the two models' posterior probabilities in light of the data, under the assumption that the two theories are equally likely to be true a priori. 

    We sample posteriors and compute evidences for two classes of models for each planet pair's RV data. 
    The first model class is a conventional $N$-body model that computes the RV signal by integrating the equations of motion governing the planetary system.
    The $N$-body model's parameters include the planets' masses $m_i$ and orbital parameters at a reference epoch, chosen to be the median time of the RV observations.
    We parameterize the planets' orbits in terms of their orbital periods $P_i$, mean anomalies $M_{i}$, eccentricities $e_i$, and arguments of periapse $\omega_i$.
    Our model's prior probabilities are uniform in the planets' orbital parameters 
    and log uniform in the planets' masses.
    {
        We adopt the usual strategy of sampling in variables $\sqrt{e_i}\cos\omega_i$ and $\sqrt{e_i}\sin\omega_i$ in our MCMC fitting to improve convergence behavior. 
        The nested sampling algorithm that we use requires variables be normalizable, so, in addition to restricting $e_i\in [0,1)$ and restricting angular variables to lie between 0 and $2\pi$, we restrict the range of our priors in the variables $P_i$ and $m_i$:
      priors on $P_i$ are uniform within $\pm25\%$ of values determined by a maximum likelihood fit and priors on $m_i$ are log uniform within a factor of 1/5 and 5 times the maximum likelihood fit values.\footnote{
      {These prior ranges in period and planet mass are significantly larger than the ranges over which the posterior has non-negligible probability density, as determined from our MCMC fits. In numerical tests we found that extending the  prior ranges for planet periods and masses led to moderately higher Bayes factors in favor of the ACR model described below.
      }
     }
    }
    In addition to the planets' orbital parameters, the model includes an RV zero point, $\gamma_k$, and instrumental jitter term, $\sigma_k$, for each set of RV observations.
    We adopt uniform priors for the zero points, $\gamma_k$, {between the minimum and maximum values of each instrument's reported radial velocity measurements.}
    We adopt log uniform priors for the jitter terms, $\sigma_k$, {ranging between ${10}^{-3}$ and $30$ times each instrument's median reported uncertainty.}
    Parameters in the $N$-body model that lead to close encounters within 3 planet Hill radii or less during the integration are automatically rejected.
    For simplicity, we fix the host star mass to the reported best-fit value from the literature and assume the planets' orbits are coplanar with an inclination of $90^{\circ}$, i.e., the system is observed edge on.
    We will refer to this model as the ``unrestricted" model.

    The second model class presumes that the RV data are the product of a pair of planets residing in an ACR configuration. 
    This model is similar to the RV model developed by \citet{Baluev2008}.
    Restricting planets to an ACR configuration reduces the degrees of freedom of the model. 
    We parameterize an ACR model for a given $j$:$j-k$ resonance in terms of the inner planet's  RV semiamplitude, $K_1$, period $P_1$, time of transit $T_{c,1}$, and argument of periapse $\omega_1$, along with $m_2/m_1$, the mass ratio of the outer planet relative to the inner, and a parameter, $s$, described below that sets the normalized AMD of the planet pair.
    Our ACR models also include the same instrumental parameters, $\gamma_k$ and $\sigma_k$, used in the unrestricted model, {and we adopt the same priors for these parameters in the ACR model as in the unrestricted model}.
    The parameter $s \in [0,1]$ sets the AMD of a planet pair to be ${\cal D} = s^2\ {\cal D}_\text{max}$ where ${\cal D}_\text{max}$ is the AMD at the point where all the ACR tracks for different planet-mass ratios intersect at a global maximum of the averaged disturbing function \citep[see][]{Michtchenko2006}.\footnote{While ACR tracks generally extend to AMDs greater than ${\cal D}_\text{max}$, we find that, in practice, the posterior density of the systems we fit is already negligible at smaller eccentricities.} 
    {
    Because, at small to moderate eccentricity values,
    ACR configurations obey $\beta_1 \sqrt{\alpha} e_1 \propto \beta_2 e_2$ \citep[e.g.,][]{Deck2015} this definition of $s$ implies that $e_i\propto s$ for small to moderate eccentricities.}
    In our ACR model MCMC fits, we sample in the variables $\sqrt{s}\cos\omega_1$ and $\sqrt{s}\sin\omega_1$ rather than $s$ and $\omega_1$. Consequently, our priors are uniform in $s$ and $\omega_1$.
    We adopt uniform priors in $P_1$ and $T_{c,1}$.
    The orbital period of the outer planet's RV signal is set to $P_2=\frac{j}{j-k}P_1$, and the semiamplitude of the outer planet is given by
 \begin{equation*}
     K_2 = \frac{m_2}{m_1}\fracbrac{j-k}{j}^{1/3}\frac{\sqrt{1-e_1^2}}{\sqrt{1-e_2^2}}K_1~.
 \end{equation*}
     The argument of periapse of the outer planet is set to $\omega_2=\omega_1+\pi$.
    The ACR model includes, in addition to the continuous parameters just described, a discrete degree of freedom that fixes the relative orbital phases of the planet pair. 
    Specifically, with the resonant angle $\sigma_1$ fixed at its equilibrium value $\sigma_1=0$ (when $k$ is odd) or $\sigma_1=\pi/k$ (when $k$ is even), the orbital phase of the outer planet is related to the orbital phase of the inner planet by 
 \begin{equation}
     M_2 = \left(1-\frac{k}{j}\right)M_1 +\frac{1-j+k}{j}\pi  + n\frac{2\pi}{j}
     \label{eq:rv:methods:discrete_dof_defn}
 \end{equation}
    for $n=0,1,...j-1$.
    We handle this discrete degree of freedom by running multiple nested sampling fits, each with $n$ fixed to one of the $j$ possible values. 
    Then, assuming that any of the $j$ possible values are equally likely \emph{a priori}, the evidence of the ACR model computed according to 
    \begin{equation}
        p(D|{\cal M}_\mathrm{ACR}) = 
        \frac{1}{j}\sum_{m=0}^{j-1} p(D|{\cal M}_\mathrm{ACR},n=m)
        \label{eq:rv:methods:discrete_dof}
    \end{equation}
    where $p(D|{\cal M}_\mathrm{ACR},n=m)$ is the evidence computed with $n$ in Equation \eqref{eq:rv:methods:discrete_dof_defn} fixed to $m$.

    Likelihoods for both models are computed in the standard way from a $\chi^2$ value that includes jitter terms added in quadrature to the reported measurement uncertainties.
    We use the \texttt{radvel} \citep{radvel} code, modified to include our $N$-body and ACR models, to forward-model RVs, evaluate model likelihoods, and run MCMC simulations. 
    We use the nested sampling code \texttt{dynesty} \citep{Speagle2020} to estimate the Bayesian evidence of each model.
    Specifically, we use the \texttt{dynesty.NestedSampler} class with default settings.
    For the ACR model, RV signals are synthesized as the sum of two planets' Keplerian RV signals rather than by means of $N$-body simulation. 
    We discuss the validity of this approximation in Appendix \ref{app:validity} where we demonstrate that it has negligible impact over the timescales spanned by the observations we fit.
    %

%%%%%%%%%%%%%%%%%%%%%%%%%%%%%%%%%%%%%%%%%%%%%%%%%%%
%%%%%%%%%%%%%%%%% Fit Results Methods %%%%%%%%%%%%%
%%%%%%%%%%%%%%%%%%%%%%%%%%%%%%%%%%%%%%%%%%%%%%%%%%%
\subsection{Results}
\label{sec:rv:data}
In this section we fit and compare our $N$-body and ACR models to two systems hosting potentially resonant giant-planet pairs. 
%%%%%%%%%%%%%%%%%% HD 45364 %%%%%%%%%%%%%%%%%%%
\subsubsection{HD 45364}
\label{sec:rv:data:HD45364}
\begin{figure*}
    \centering
    \includegraphics[width=\textwidth]{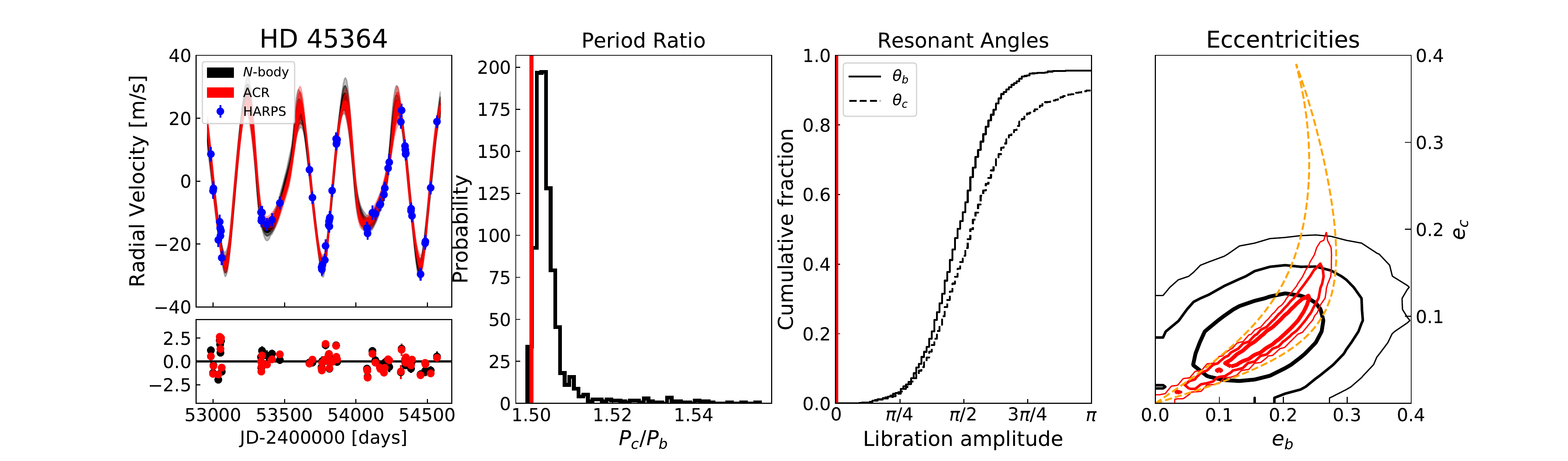}
    \par
    \vspace*{\floatsep}
  \begin{minipage}{\textwidth}
    \centering
    \begin{tabular}{|c| c c c | c c c|}
        \hline
        \input{HD45364_fit_summary.tex}
        \hline
    \end{tabular}
    \caption{
    Summary of fit results for HD 45364.
    { Top:} comparison of  unrestricted model fits against 3:2 ACR model fits for HD 45364.
    For all plots, $N$-body model results are plotted in black, 
    while the ACR model results are plotted in red. 
    The leftmost panel shows the RV solutions derived from both models, using shaded regions to illustrate the $1\sigma,2\sigma$, and $3\sigma$ distributions for each. 
    The bottom of the panel shows the normalized residuals of each model,
    i.e., the model residuals divided by the model error bars including the allowance for jitter. Points and error bars indicate the median and 1$\sigma$ range of the normalized residuals.
    The middle-left panel shows the distribution of average period ratios derived from both models.
    The middle-right panel shows the cumulative distribution of resonant angle libration amplitudes measured via $N$-body integrations.
    The rightmost panel shows the eccentricity posteriors for the two planets.
    The dashed orange lines in the right-hand plot indicate the ACR tracks that encompass the full range of the planet mass ratios in the ACR model posterior sample.
    {Bottom:} table giving log evidences from nested sampling simulations and 
    percentiles of parameters' marginalized posterior distributions from MCMC.
    }
    \label{tab:HD45364_fit_summary}
  \end{minipage}
\end{figure*}%
    The first system we fit is HD 45364, a system of two planets in a 3:2 MMR orbiting a K0 star of mass $M_*=0.82~M_\odot$
    originally discovered by \citet{Correia2009}.
    The best-fitting solution reported by \citet{Correia2009} displays large resonant angle libration amplitudes and secular eccentricity variations and is therefore not in an ACR configuration.
    %
    %\citet{Correia2009} also show that, given the planets' masses, much of the phase space outside of the 3:2 MMR is unstable.
    %
    A number of subsequent studies have further explored the dynamics and possible formation histories of the HD 45364 system.
    \citet{Rein2010} perform a suite of hydrodynamical simulations of planet--disk interactions in order to explore potential migration histories leading to the present-day orbital configuration of HD 45364 system. 
    Their simulations produce systems with orbital parameters significantly different from the \citet{Correia2009} best-fit solution but with RV signals that fit the observations with approximately equal statistical significance.
    \citet{Correa-Otto2013} show that the dynamical configurations similar to those found by both \citet{Correia2009} and \citet{Rein2010} can be obtained through simple parameterized migration simulations with exponential migration and eccentricity-damping forces.
    As expected, the planets in the simulations of \citet{Correa-Otto2013} are driven into ACR configurations.
    While \citet{Correa-Otto2013} note that the ACR configurations' lack of libration amplitude is at odds with the orbital solutions reported by \citet{Correia2009} and \citet{Rein2010}, they do not attempt to determine whether the RVs of HD 45364 can be adequately fit with planets in a zero libration amplitude configuration.
    \citet{Delisle2015} also analyze the HD 45364 system in order to constrain properties of the system's migration during formation, similar to Section \ref{sec:discussion:migration} below. 
    However, their constraints are derived under the assumption that the system possesses the nonzero libration amplitude measured by \citet{Correia2009} and they do not explore the possibility that the system may in fact reside in an ACR configuration.

    Figure \ref{tab:HD45364_fit_summary} summarizes our fit results for HD 45364.
    The ACR model is formally preferred by our nested sampling simulations, with a log Bayes factor of {$\sim 7$}.
    The rightmost panel of Figure \ref{tab:HD45364_fit_summary} shows that the posterior distribution of the $N$-body model planet eccentricities  are concentrated quite close to the ACR values.
    We integrate 1500 initial conditions randomly selected from the unrestricted model MCMC posterior sample for 100 orbits of planet b in order to determine the resonant angle libration amplitudes and average period ratios plotted in the middle panels of Figure \ref{tab:HD45364_fit_summary}.
    Resonant angle libration amplitudes were recorded by determining the maximum deviations of the angles $\theta_b =3\lambda_c-2\lambda_b-\varpi_b$ and  $\theta_c=3\lambda_c-2\lambda_b-\varpi_c$ from their respective equilibrium values of $0$ and $\pi$ over the course of the integration.
    Cumulative distributions of libration amplitudes are plotted in the center-right panel of Figure \ref{tab:HD45364_fit_summary}.
    While the majority of initial conditions show resonant libration for both angles with amplitudes $\lesssim \pi/2$, neither cumulative distribution reaches 1 due to a small fraction of simulations in which one or both resonant angles circulate.
    The moderate to large libration amplitudes inferred when fitting the unrestricted model may initially appear to be at odds with the preference for the zero libration amplitude ACR model inferred from the Bayesian evidence calculations.  
    This apparent contradiction suggests that  the inference of large libration amplitudes in the unrestricted model is principally driven by the larger phase-space volume occupied by these configurations and not by any significant improvement in fit derived from large libration amplitude solutions.
    This example serves to illustrate that the influence of priors must be carefully considered when deriving conclusions about the dynamics of planetary systems from RV fits.

% %%%%%%%%%%%%%%%%%% HD 33844 %%%%%%%%%%%%%%%%%%
\subsubsection{HD 33844}
\label{sec:rv:data:HD33844}
\begin{figure*}
\label{fig:HD33844rv_combo_plot}
    \centering
    \includegraphics[width=\textwidth]{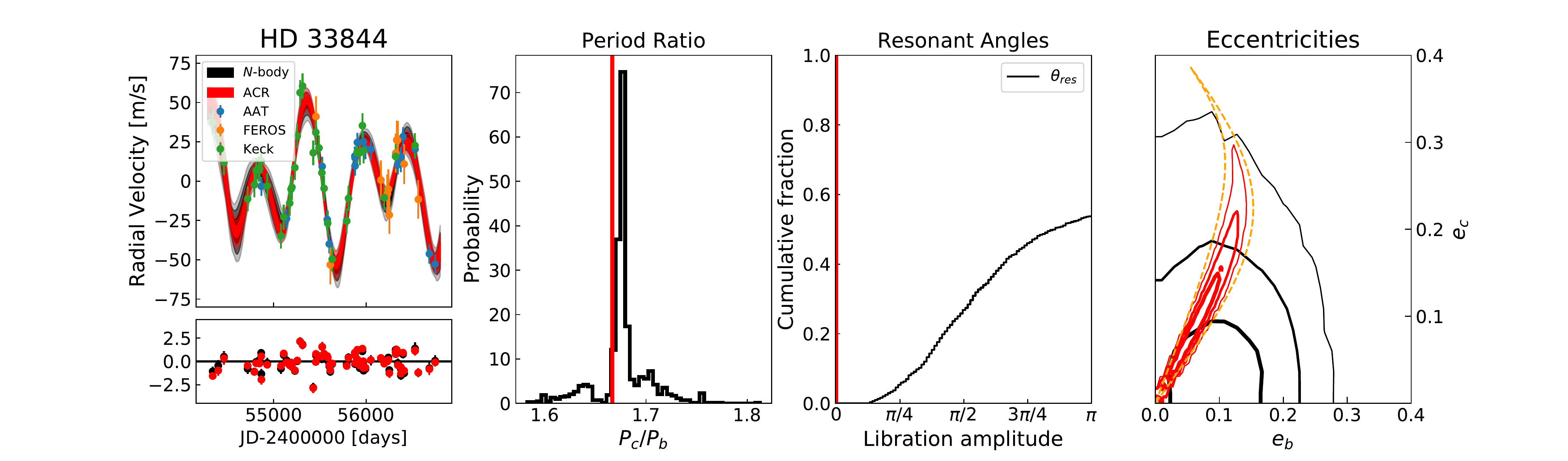}
    \par
    \vspace*{\floatsep}
  \begin{minipage}{\textwidth}
    \centering
    \begin{tabular}{|c| c c c | c c c|}
        \hline
        \input{HD33844_fit_summary.tex}
        \hline
    \end{tabular}
    \caption{
    Summary of fit results for HD 33844. 
    {Top:} 
    comparison of unrestricted model fits against 5:3 ACR model fits for HD 33844.
    For all plots, $N$-body model results are plotted in black, 
    while the ACR model results are plotted in red. 
    The leftmost panel shows the RV solutions derived from both models, using shaded regions to illustrate the $1\sigma,2\sigma$, and $3\sigma$ distributions for each. 
    The bottom of the panel shows the normalized residuals of each model,
    i.e., the model residuals divided by the model error bars including the allowance for jitter. Points and error bars indicate the median and 1$\sigma$ range of the normalized residuals.
    The middle-left panel shows the distribution of average period ratios derived from both models.
    The middle-right panel shows the cumulative distribution of resonant angle libration amplitudes measured via $N$-body integrations.
    The rightmost panel shows the eccentricity posteriors for the two planets.
    The dashed orange lines in the right-hand plot indicate the ACR tracks that encompass the full range of the planet mass ratios in the posterior sample.
    {Bottom:}
    table giving log evidences from nested sampling and 
    the percentiles of parameters' marginalized posterior distributions from MCMC.
    }
    \label{tab:HD33844_fit_summary}
  \end{minipage}
\end{figure*}%
The second system we fit, HD 33844, consists of two giant planets orbiting in or near a 5:3 MMR and was originally reported by  \citet{Wittenmyer2016}. 
We adopt the stellar mass of $M_* = 1.84~M_\odot$ for HD 33844 based on the best-fit value reported by \citet{Stassun2017}.
Combining RV fits with longer-term $N$-body simulations to rule out RV solutions leading to dynamical instabilities on timescales less than 100 Myr, \citet{Wittenmyer2016} find that many of the stable orbital configurations in the neighborhood of their best-fit solution reside in the 5:3 MMR.\footnote{
    \citet{Wittenmyer2016} find that the angle $5\lambda_c - 3\lambda_b - \omega_b + \omega_c$ alternates between libration and circulation in their simulations. 
    This combination, however, is not a dynamically meaningful resonant angle as it is not invariant under rotational transformations of the coordinate system.
 } 

Figure \ref{tab:HD33844_fit_summary} summarize our fit results for HD 33844.
Our inferred planet masses and orbital parameters are in good agreement with values reported by \citet{Wittenmyer2016}.
The unrestricted and ACR models provide similar quality fits to the data and, due to its lower dimensionality, the ACR model is formally preferred with a log Bayes factor of {$\sim 6$} based on the results of our nested sampling simulations.

In order to more clearly interpret the resonant state of 
RV solutions inferred with the unrestricted model, we turn to the analytic resonance model of \citet{Hadden2019}.
While the full dynamics of a 5:3 MMR are captured by a two-degree-of-freedom system with two resonant angles, $\sigma_1$ and $\sigma_2$ defined in Section \ref{sec:capture}, \citet{Hadden2019} shows that, to an excellent approximation,  the dynamics of the MMR only depend on a single resonant angle $\theta_\text{res} = 5\lambda_c - 3\lambda_b - 2z$, where $z\approx \text{arg}(e_c e^{i\varpi_c}-e_b e^{i\varpi_b})$.
To determine the resonant behavior of the RV solutions fit with the unrestricted model, we again integrate 1500 initial conditions randomly selected from the MCMC posterior sample for 100 orbits of planet b and record angle libration amplitudes as the maximum deviation of the angle $\theta_\text{res}$ from its equilibrium value, $\pi$.
(ACR solutions have periodic variations in $\theta_\text{res}$ on the synodic timescale about the equilibrium of the averaged system with an amplitude of approximately $\sim 0.1\pi$.)
Approximately $50\%$ of these initial conditions show librating behavior, and the cumulative distribution of their libration amplitudes are recorded in the middle-right panel of Figure \ref{tab:HD33844_fit_summary}.
The rightmost panel shows that the $N$-body model prefers a larger eccentricity for planet b when the system is not restricted to lie inside the ACR. 
%%%%%%%%%%%%%%%%%%%%%%%%%%%%%%%%%%%%%%%%%%%%%%%%
%%%%%%%%%%%%%%%%%%%%% Discussion %%%%%%%%%%%%%%%
%%%%%%%%%%%%%%%%%%%%%%%%%%%%%%%%%%%%%%%%%%%%%%%%
\section{Discussion}
\label{sec:discussion}
\subsection{Dynamical stability}
\label{sec:discussion:stability}
    \begin{figure*}
    \centering
    \includegraphics[width=0.4\textwidth]{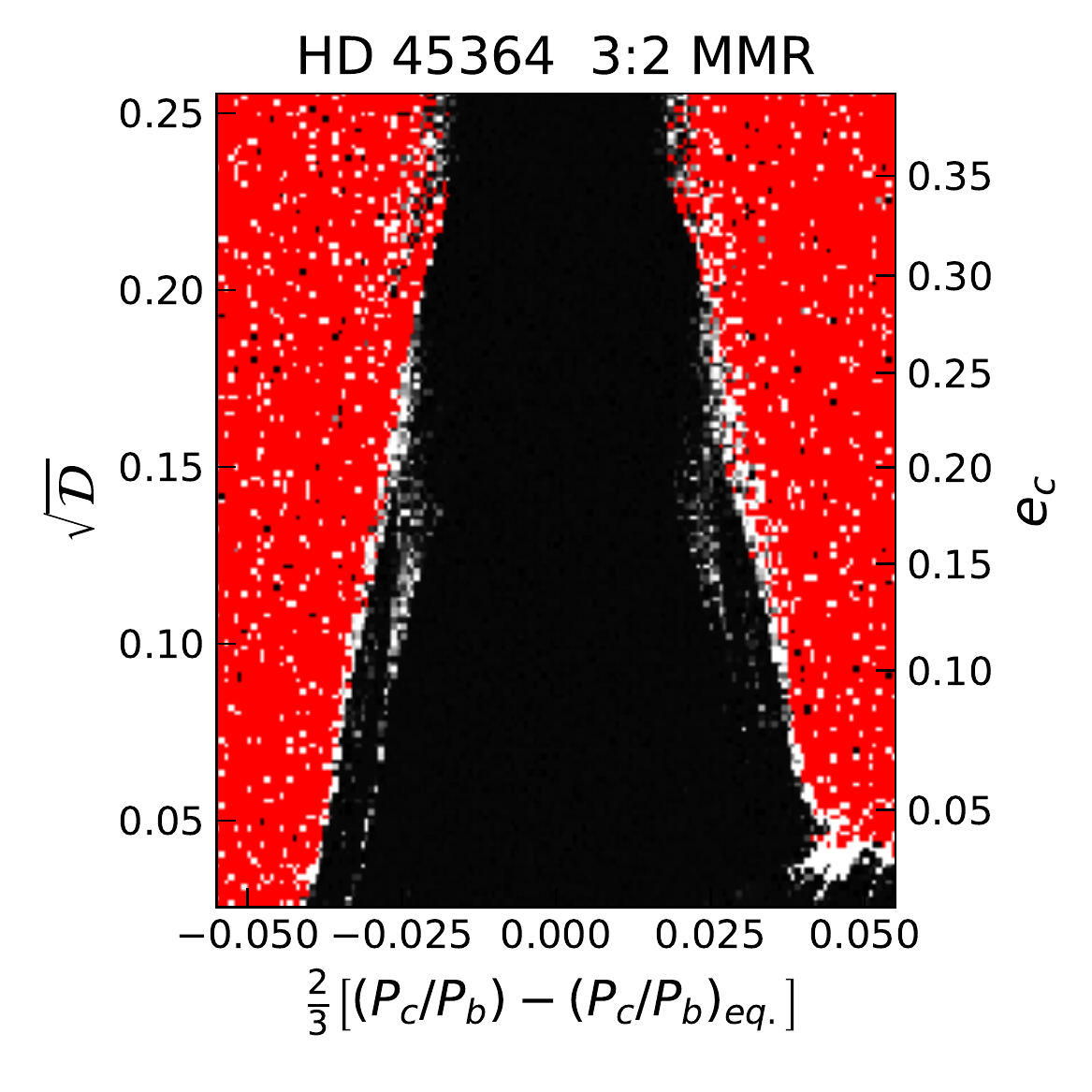}
    \includegraphics[width=0.4\textwidth]{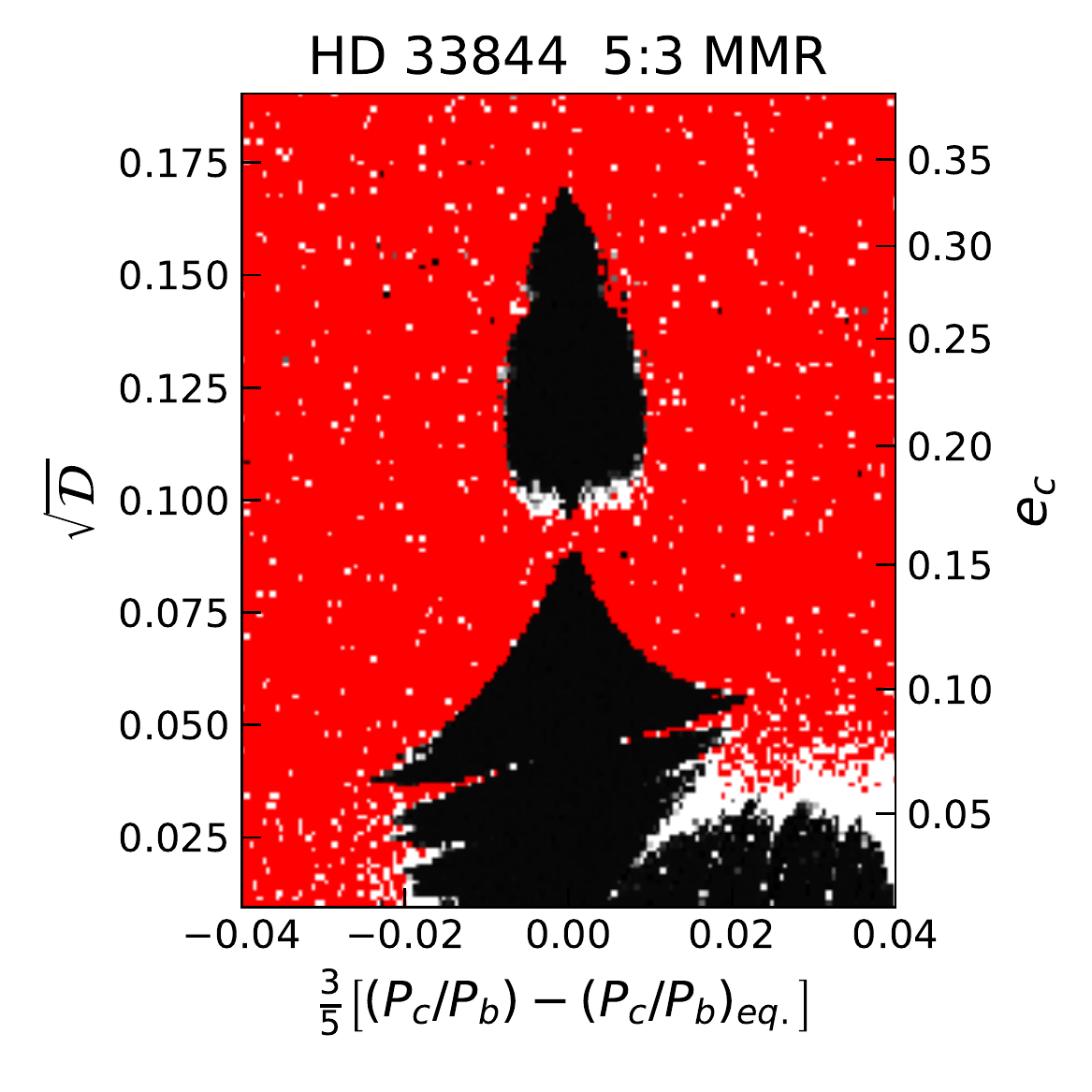}
    \caption{
    Stability maps in the vicinity of the 3:2 ACR of HD 45364 (left) and the 5:3 ACR of HD 33844 (right), respectively. 
    Planet masses are set to their median values from the ACR model MCMC.
    Each panel shows results from a grid of simulations run for 3000 orbits of the outer planet.
    Black indicates regular trajectories, 
    white indicates chaotic trajectories,
    and red indicates trajectories that
    experienced a close encounter during the simulation.
    We use a grayscale stretching from MEGNO values of 1.9 to 5 in order to color regular and chaotic points.
    }
    \label{fig:systems:stability}
\end{figure*}

    So far, we have not considered the long-term stability of planets' orbital configurations in our fits. 
    The high planet masses and close spacings of the HD 45364 and HD 33844 systems place them near the boundary of dynamical instability.
    This dynamical fragility can be leveraged to derive tighter constraints on the planets' masses and orbital properties by rejecting any orbital configurations that rapidly go unstable \citep[rejection sampling based on long-term stability is common practice in the literature: e.g.,][]{Wittenmyer2016}.
    Because we are primarily interested in evaluating whether systems' RVs can be explained by planets in an ACR configuration rather than deriving precise orbital constraints, we will not apply this rejection sampling procedure here.
    Instead, we explore the dynamical stability in the vicinity of ACR configurations by means of the dynamical maps shown in Figure \ref{fig:systems:stability}.
    The maps in Figure \ref{fig:systems:stability} indicate dynamical stability results for a grid of short $N$-body integrations.
    In each integration, planets are initialized with masses equal to the median planet masses derived from our MCMC fits and orbital elements corresponding to an ACR equilibrium, except that we vary the planets' period ratio away from the equilibrium value.
    We explore a range of planet eccentricities by running integrations over a grid of normalized AMD values up to the value ${\cal D}_\text{max}$ described in Section \ref{sec:rv:methods}.
    {
    Each integration is initialized with $\lambda_2=\lambda_1$, and all osculating orbital elements are set to their ACR values, except for the planets' period ratios, which we vary along the $x$-axes of the maps.
    The planets' osculating elements in the ACR configuration are computed using the transformation from mean to osculating elements described in Appendix \ref{app:Equations:transformation}. 
    The simulations are run using the \texttt{WHFast} integrator \citep{WHFast} with time steps set to $1/30$ of the periapse passage timescale of the inner planet, $T_\text{peri} = 2\pi/\dot{f}_\text{max}$, where $\dot{f}_\text{max}$ is the time derivative of the true anomaly at pericenter \citep{Wisdom2015}.
    Regular and chaotic trajectories are identified based on the MEGNO chaos indicator \citep{Cincotta2003}.
    }
   
    From Figure \ref{fig:systems:stability}, we see that, in the case of HD 45364, the 3:2 ACR provides a large island of stability in a phase space that is otherwise unstable except for a small pocket of stability just wide of the resonance at low eccentricity.
    This is in agreement with the original analysis of \citet{Correia2009}. 
    The structure of the 5:3 ACR in the HD 33844 system, 
    on the other hand, is somewhat more complicated.
    The ACR shows two separate islands of stability separated by a thin chaotic region near ${\cal D}\approx 0.1$ along with a small pocket of stability wide of the resonance at low eccentricity.

    Irrespective of whether or not the HD 45364 and HD 33844 systems are truly in ACR configurations we can conclude that they most likely reside in resonance
    since the local phase space outside of these resonances is almost entirely chaotic and unstable.
    The stability maps shown in Figure \ref{fig:systems:stability} highlight how knowledge of ACR structure allows much more efficient identification of stable parameter space than a brute-force grid search of the high-dimensional parameter space of planets' orbital elements  when fitting RVs.
    It is well known that MMRs can provide phase protection from close encounters that would otherwise destabilize strongly interacting planets.
    Because ACR configurations are elliptic equilibria of the averaged dynamical system describing the planets' resonant motion, they should be the most dynamically stable resonant configurations possible: 
    the degree of nonlinearity in the neighborhood of an elliptic equilibrium, and thus the potential for chaos and dynamical instability, increases with distance from the equilibrium  \citep{Giorgilli1989}.
    Therefore, vicinities of ACR configurations provide good starting points for any search for dynamically stable orbital configurations when modeling strongly interacting planet pairs.
    In their analysis of the HD 82943 system,
    \citet{Baluev2014} previously noted the utility of enforcing ACR configurations to ensure dynamical stability when fitting RV data.
    Conversely, if no stable ACR configuration exists for a given set of planet masses, then stable orbital configurations likely do not exist in the nearby phase space.

%%%%%%%%%%%%%%%%%%%%%%%%%%%%%%%%%%%%%%%%%%%%
\subsection{Inferring Migration History}
\label{sec:discussion:migration}
\begin{figure*}
    \centering
    \includegraphics[width=0.45\textwidth]{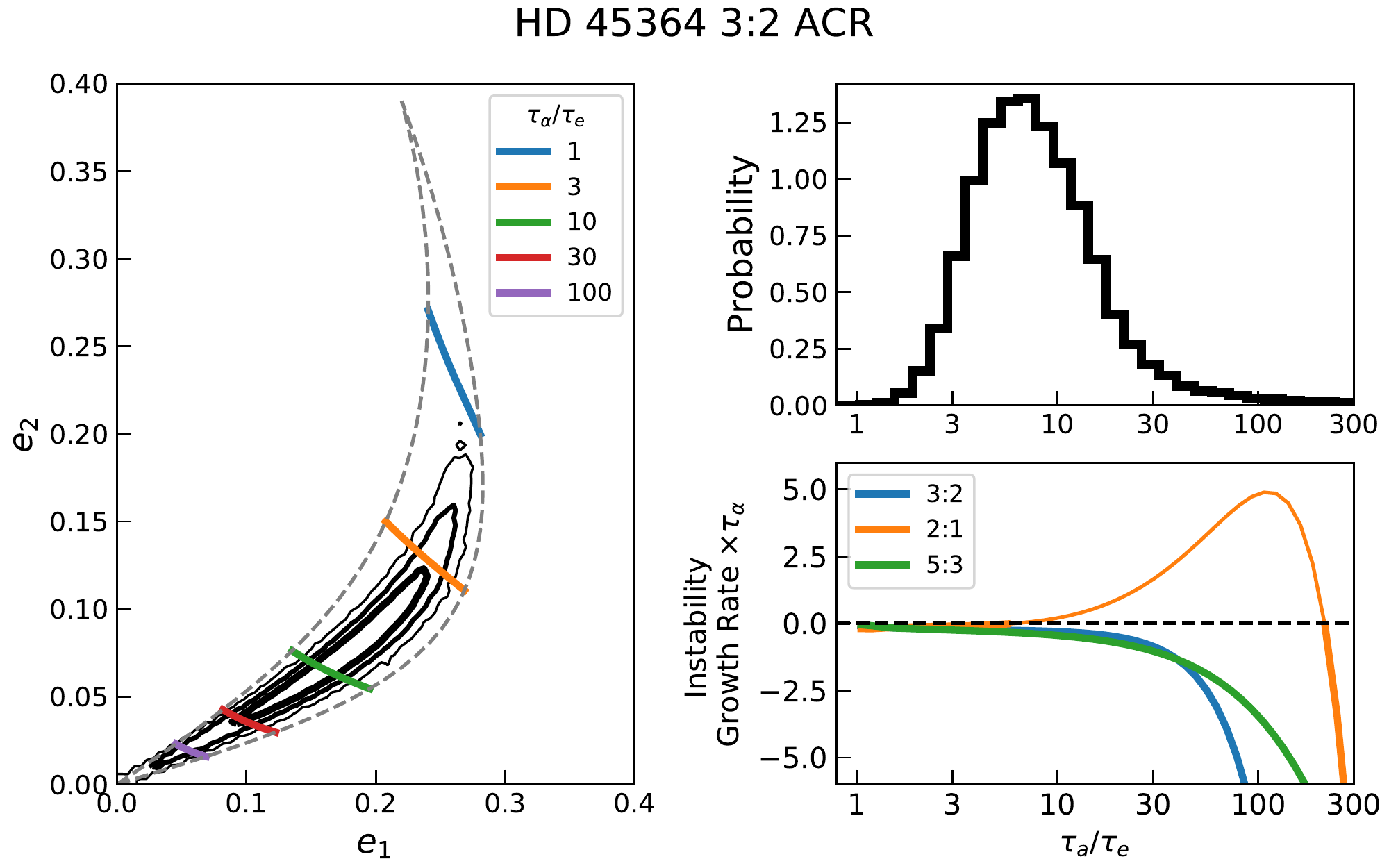}
    \includegraphics[width=0.45\textwidth]{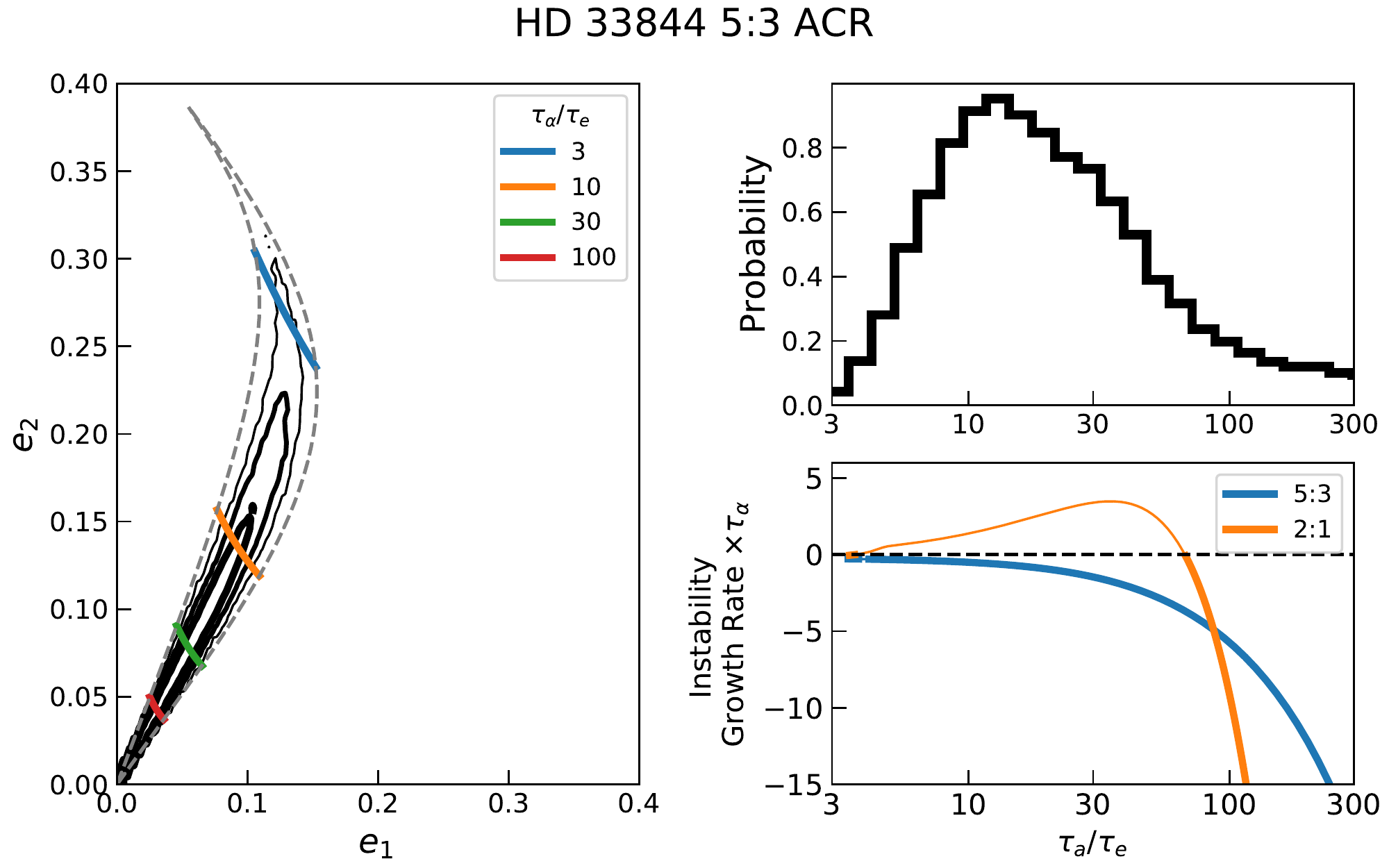}
    \caption{
    Inferring the relative strength of migration and eccentricity damping, $\tau_\alpha/\tau_e$, during resonant capture for  HD 45364 (left) and HD 33844 (right) from our ACR model fits.
    For each system, the left panel shows the posterior distribution of planet eccentricities from the ACR model MCMC fit with the loci of equilibrium configurations for different values of $\tau_\alpha/\tau_e$ plotted in different colors.
    The upper-right panels show the posterior distribution of $\tau_\alpha/\tau_e$ values inferred by converting the eccentricity values shown in the left panels to $\tau_\alpha/\tau_e$ values.
    The bottom-right panel shows the  stability/instability growth rates, in units of $\tau_\alpha^{-1}$, as a function of $\tau_\alpha/\tau_e$ for the current resonance occupied by each system as well as the 2:1 MMR and, for HD 45364, the 5:3 MMR.
    Positive values correspond to instability (the ACR becomes an unstable spiral in the presence of dissipation) 
    while negative values correspond to stability (the ACR becomes a stable spiral in the presence of dissipation).
    }
    \label{fig:HD45364.migration_summary}
\end{figure*}
% ACR configuration is an equilibrium  
    {
    In this section we explore how the present-day dynamical configurations can be used to glean information about planet--disk interactions that presumably established these configurations during the planet formation process. Because theoretical understanding of planet migration is incomplete, we rely on our simple parameterized model of eccentricity damping and migration in Equation \eqref{eq:capture:dissipation_forces}. 
    While the migration and eccentricity damping rates experienced by planets may vary over the course of the formation process in ways that are not captured by our simple parameterized damping model, the model is still useful for capturing qualitative aspects of planet--disk interactions during the formation process.
    The quantitative constraints derived with our simple migration model provide useful metrics against which predictions of more rigorous theoretical and numerical studies of planet migration can be compared. 
    }

    If it is assumed that planet pairs were subject to migration and eccentricity-damping forces for a time period sufficiently long to reach dynamical equilibrium, a given ACR equilibrium configuration corresponds directly to a ratio of migration timescale to eccentricity-damping timescale {at the time the equilibrium was established}.
    This is illustrated in Figure \ref{fig:HD45364.migration_summary}. 
    The figure shows the posterior distributions of each planet pair's eccentricities from our MCMC fits with ACR models. 
    In different colored lines, we plot the loci of equilibria along these ACR tracks for different values of $\tau_\alpha / \tau_e$ where we define $\tau_e^{-1} \equiv \tau_{e,1}^{-1} +\tau_{e,2}^{-1}$ and assume that $m_2\tau_{e,2} = m_1\tau_{e,2}$ so that the strength of eccentricity damping experienced by a planet is proportional to its mass. 
    This simple prescription approximately matches the expected scaling of eccentricity damping experienced due to type I migration \citep{Kley2012} or dynamical friction from interactions with planetesimals \citep{Ida1990}.
    For each system, Figure \ref{fig:HD45364.migration_summary} also shows the distribution of  $\tau_\alpha / \tau_e$ values inferred by mapping our posterior samples of mass ratios and eccentricities to values of  $\tau_\alpha / \tau_e$. For HD 45364 we infer  $\tau_\alpha / \tau_e=8_{-3}^{+8}$ and for HD 33844 we find $\tau_\alpha / \tau_e=20_{-12}^{+65}$.

    {Of course, the planet pairs could have failed to reach an equilibrium between migration and eccentricity-damping forces and instead have been stranded at their current resonant configurations 
    if the gas disk dissipated before any equilibrium was reached.
    In this case, the posterior distributions of $\tau_\alpha/\tau_e$ derived from our ACR model fits would serve as lower limits on the timescale of eccentricity damping relative to the migration rate: had the eccentricity damping been stronger, the systems would have reached equilibria at lower eccentricities while ${\cal D}$ increased monotonically after capture into the resonance. Far from an equilibrium between migration and eccentricity damping forces, the eccentricity damping will play a negligible role in the evolution of ${\cal D}$ and
    \begin{eqnarray}
        \dot {\cal D} \approx \frac{\beta_1\beta_2\sqrt{\alpha}}{\beta_1\sqrt{\alpha}(1+s) + \beta_2 s}\frac{1}{2\tau_\alpha}~.\label{eq:ddot_approx}
    \end{eqnarray}
    Inserting the appropriate parameters in Equation \eqref{eq:ddot_approx}, HD 45364 would reach the median posterior value of ${\cal D}$ increasing from 0 in a time of $\sim0.15\tau_\alpha$.
    Similarly, HD 33844 would reach its median value of ${\cal D}$ in $\sim 0.03\tau_\alpha$.
    The times the two systems would have required to reach their current MMRs by large-scale migration from a period ratio of approximately $\sim$2:1, are $\sim0.2\tau_\alpha$ for HD 45364 and $\sim0.1\tau_\alpha$ for HD 33844. 
    This scenario therefore requires that planets' migration timescales to be comparable to or longer than the lifetime of the protoplanetary disk while
    migration rates predicted by conventional type II migration theory are usually at least one or two orders of magnitude shorter than disk lifetimes \citep{Kley2012}.
}

\begin{figure*}
    \centering
    \includegraphics[width=0.95\textwidth]{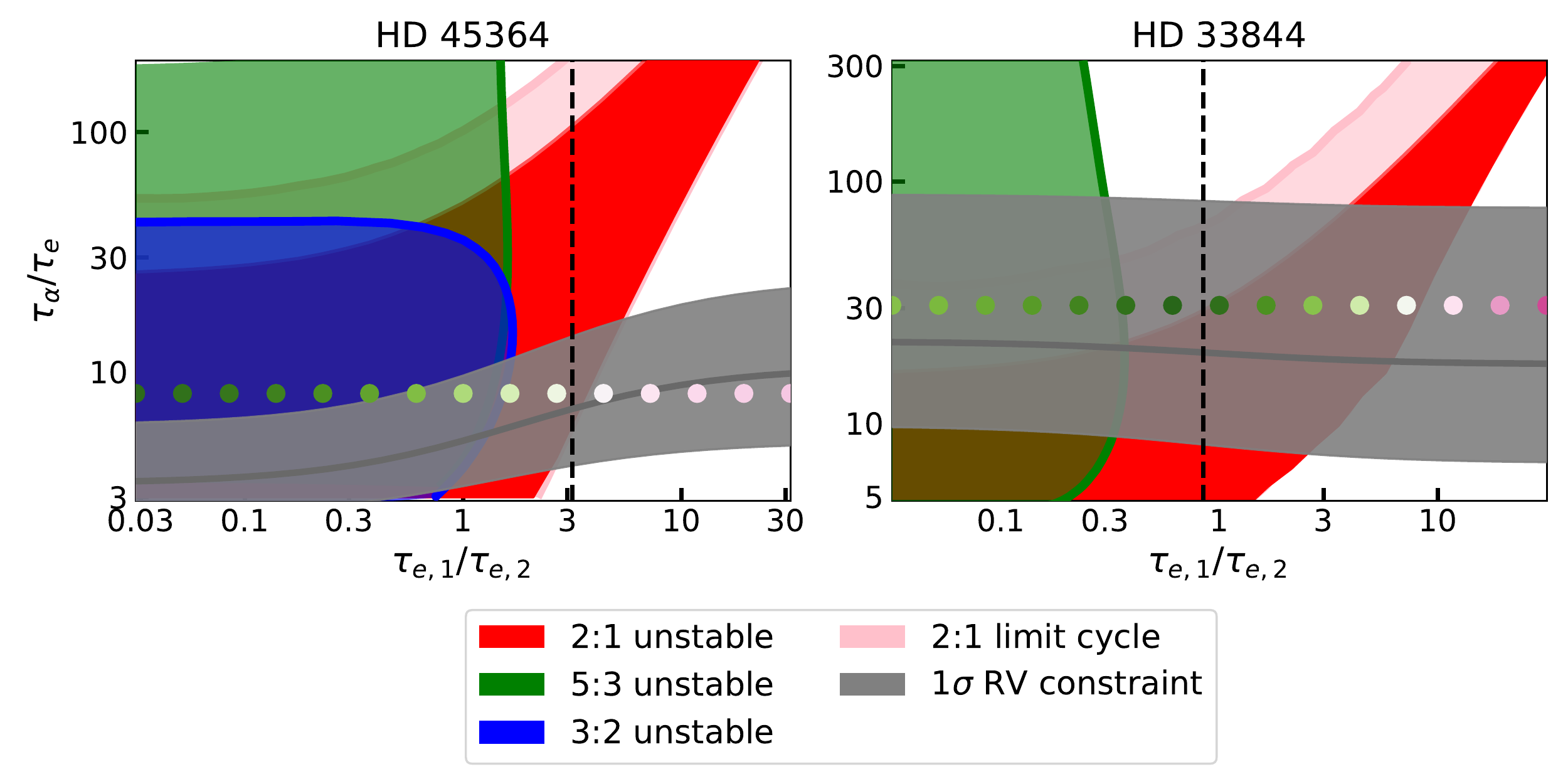}
    \caption{
    Stability of ACR equilibria as a function of
    $\tau_{\alpha}/\tau_e$ and $\tau_{e,1}/\tau_{e,2}$ for the HD 45364 and HD 33844 systems.
    Eccentricity and semi-major axis damping evolve planets toward an ACR equilibrium that may be stable or unstable, depending on the relative eccentricity damping strengths experienced by the inner and outer planet.
    Regions where these equilibria are unstable are indicated by different colors for different resonances.
    Parameter combinations that reach a limit cycle in the 2:1 MMR rather than passing through the resonance are also indicated.
    Gray bands show the  1$\sigma$ inferred ranges of $\tau_{\alpha}/\tau_e$ as a function
    of $\tau_{e,1}/\tau_{e,2}$ based on our ACR model MCMC fit.
    The vertical dashed line indicates $\tau_{e,1}/\tau_{e,2} = m_{2}/m_1$, the eccentricity damping relationship presumed in Figure \ref{fig:HD45364.migration_summary}.
    Colored circles indicate the parameters used in the $N$-body simulations plotted in Figures \ref{fig:HD45364.example_history} and \ref{fig:HD33844.example_history}.
    For both HD 45364 and HD 33844, there exist regions of parameter space that would have both led to their escape from the 2:1 MMR  while also allowing for subsequent stable capture into their current resonant configurations.
    }
    \label{fig:HD45364.acr_stability}
\end{figure*}

One of the puzzling features of the HD 45364 and HD 33844 systems is that they both lie interior to the 2:1 MMR. 
In standard core accretion theories of planet formation, massive planets like those in these systems are expected to form farther apart than the 2:1 MMR \citep[][and references therein]{armitage2007lecture}.
If the resonances in these systems were established through convergent migration, then, during formation, the planets should have first encountered and been captured into the 2:1 MMR.
One possibility is that their migration was sufficiently rapid to avoid capture in the 2:1 MMR.
This is the scenario considered, for example, in the simulations of \citet{Rein2010} to explain the orbital configuration of HD 45364.
Another possibility is that capture in the 2:1 MMR was rendered temporary in both systems because the ACR configuration was an unstable spiral in the presence of dissipative forces.
{This mechanism presents a compelling alternative explanation because, as \citet{Rein2010} note, the migration rates required to avoid capture in the 2:1 MMR are significantly faster than the rates of migration predicted by conventional type II migration theory. }
We explore this possibility further in Figures \ref{fig:HD45364.migration_summary}  
and \ref{fig:HD45364.acr_stability}.%

Figure \ref{fig:HD45364.migration_summary} illustrates the stability of ACR equilibria as a function of $\tau_\alpha/\tau_e$ for 
the current resonance occupied by each system as well as the 2:1 MMR (plus the 5:3 MMR for HD 45364 system), which the planets would have traversed on their way to their current MMR configuration. 
The curves are computed as follows: first, we determine the equilibrium configuration for a given $\tau_{\alpha}/\tau_e$ using the resonance equations of motion, taking $p=1$ in Equation \eqref{eq:capture:dissipation_forces} describing the dissipative forces experienced by the planets.
Next, we compute the Jacobian of the equations of motion at this equilibrium.
The stability is then determined based on the real parts of the eigenvalues of the Jacobian as described in Section \ref{sec:capture}.  
The plot shows the maximum real eigenvalue component in units of $\tau_{\alpha}^{-1}$, which we have taken to be $\tau_\alpha=3\times 10^5P_2$. (We confirm that, for sufficiently large $\tau_\alpha$, growth rates simply scale with $\tau_\alpha^{-1}$.)
For HD 45364, the 2:1 ACR equilibrium is unstable for $6<\tau_\alpha/\tau_e<211$. This range encompasses 65\% of the posterior points plotted in the top right panel of Figure \ref{fig:HD45364.migration_summary}.
For HD 33844, instability occurs at the 2:1 MMR for $3.6<\tau_\alpha/\tau_e<67$, encompassing 81\% of the posterior. 

In Figure \ref{fig:HD45364.acr_stability}, we perform similar calculations, but we now relax our assumption of  $\tau_{e,1}/\tau_{e,2} = m_2/m_1$ and instead explore a range of relative eccentricity-damping strengths.
We examine credible regions of $\tau_{a}/\tau_e$ inferred from our ACR model fit posterior as a function of $\tau_{e,1}/\tau_{e,2}$.
The figure also shows the regions for which different combinations of $\tau_{e,1}/\tau_{e,2}$ and $\tau_{\alpha}/\tau_e$ drive the system to an ACR configuration that is made unstable by the dissipative forces.

While the elliptic point of the ACR configuration may be rendered unstable by dissipative forces, the ultimate outcome of the instability can lead to either passage through the resonance or a limit cycle within the resonance. 
In order for the planets of the HD 45364 and HD 33844 systems to have escaped permanent capture in a 2:1 MMR, instabilities must have led to escape rather than a limit cycle.  
Using an integrable approximation for the dynamics of a first-order MMR, \cite{Deck2015} derive an approximate criterion to determining whether instability is expected to result in escape or a limit cycle.
Their criterion can be stated in terms of a critical value of ${\cal D} = {\cal D}_\text{bif}$, where the Hamiltonian  ${\bar H}({\bf z },{\cal D})$  exhibits a bifurcation and goes from having one elliptic fixed point to having two elliptic fixed points and one unstable fixed point.
If ${\cal D} > {\cal D}_\text{bif}$ then the unstable librations grow until the system eventually escapes resonance while, for  ${\cal D} < {\cal D}_\text{bif}$, the librations saturate at a finite value when the system reaches a limit cycle.
To compute ${\cal D}_\text{bif}$ for the 2:1 MMR, we use an integrable approximation for the resonant dynamics similar to the one employed by \cite{Deck2015} implemented in the \texttt{celmech} package.\footnote{\url{https://github.com/shadden/celmech}}
Damping parameter combinations expected to lead to a limit cycle in the 2:1 MMR, where the equilibrium configuration is unstable but ${\cal D} < {\cal D}_\text{bif}$,  are indicated by pink shading in Figure \ref{fig:HD45364.acr_stability}.

\begin{figure}
    \centering
    \includegraphics[width=0.95\columnwidth]{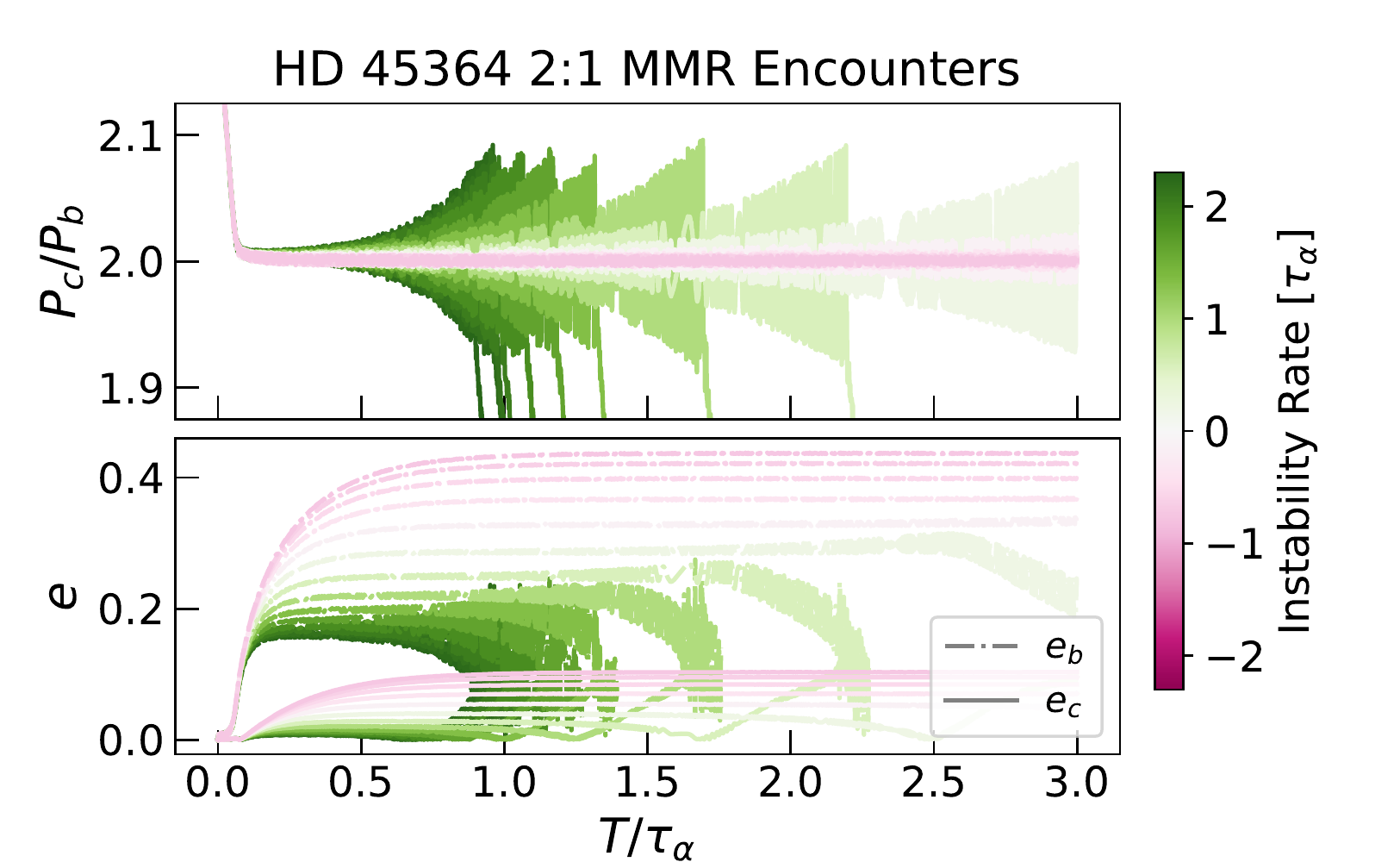}
    \caption{
    Example outcomes of a 2:1 MMR encounter for HD 45364 for different values of $\tau_{e,1}/\tau_{e,2}$ indicated by the colored points plotted in Figure \ref{fig:HD45364.acr_stability}.  The upper panel shows the planets' period ratio as a function of time measured in units of $\tau_\alpha$, and the bottom panel shows the evolution of the planets' eccentricities. Results of each simulation are colored according to the instability growth rate computed using the resonance equations of motion described in Section \ref{sec:capture}. Green colors correspond to unstable parameters and escape from the resonance after capture, while pink simulations are stable and remain in the resonance.
    }
    \label{fig:HD45364.example_history}
\end{figure}

\begin{figure}
    \centering
    \includegraphics[width=0.95\columnwidth]{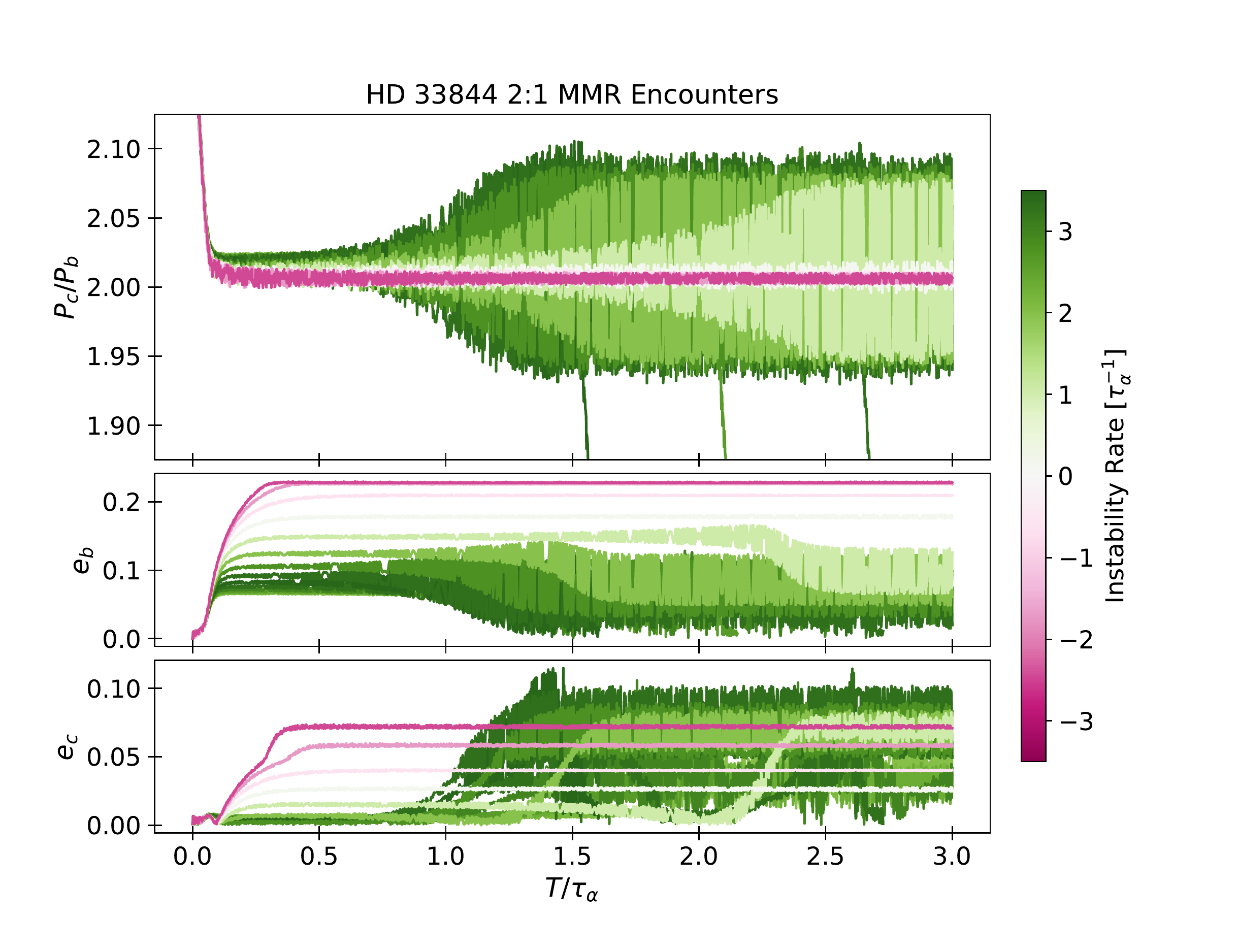}
    \caption{
    Example outcomes of a 2:1 MMR encounter for HD 33844 for different values of $\tau_{e,1}/\tau_{e,2}$ indicated by the colored points plotted in Figure \ref{fig:HD45364.acr_stability}.
    The upper panel shows the planets' period ratio as a function of time measured in units of $\tau_\alpha$, and the bottom panels show the evolution of each planet's eccentricity. 
    Results of each simulation are colored according to the instability growth rate computed using the resonance equations of motion described in Section \ref{sec:capture}. 
    Green corresponds to unstable parameters while pink simulations are stable. 
    Some of the unstable simulations pass through the resonance whereas others reach a limit cycle with large libration amplitude inside the 2:1 MMR.
    }
    \label{fig:HD33844.example_history}
\end{figure}

Figures \ref{fig:HD45364.example_history} and \ref{fig:HD33844.example_history} show example encounters with the 2:1 MMR  for HD 45364 and HD 33844, respectively, at fixed $\tau_\alpha/\tau_e$ and for a range of $\tau_{e,1}/\tau_{e,2}$ values that are indicated by the colored points in Figure \ref{fig:HD45364.acr_stability}.
The simulations in Figures \ref{fig:HD45364.example_history} and \ref{fig:HD33844.example_history} use the symplectic \texttt{WHFast} integrator, instead of the \texttt{IAS15} integrator, with time steps set to 1/100 of the initial orbital period of the inner planet.
We choose the planets' individual migration timescales such that $\tau_\alpha = 3\times10^5 P_2$ and  $\frac{m_1}{a_{1,0}\tau_{m,1}}+\frac{m_2}{a_{2,0}\tau_{m,2}}=0$. 
The latter condition ensures that the time derivative of the systems' energy is approximately zero and minimizes any inward drift of the system to ensure that dynamics remain well resolved by the simulations' fixed time steps.
Trajectories are colored according to the predicted stability of their 2:1 ACR equilibrium:
pink trajectories are stable while green trajectories are unstable. 
In both Figures \ref{fig:HD45364.example_history} and \ref{fig:HD33844.example_history} the planets are initialized with a period ratio of 2.2 on circular orbits and migrate with $\tau_\alpha = 3\times 10^5 P_2$.
For HD 45364 shown in Figure \ref{fig:HD45364.example_history},  we set $\tau_e =1/(\tau_{e,1}^{-1}+\tau_{e,2}^{-1})=8.13\tau_\alpha$ while in for HD 33844 shown in Figure \ref{fig:HD33844.example_history}, we set $\tau_e =1/(\tau_{e,1}^{-1}+\tau_{e,2}^{-1})=31\tau_\alpha$.

For HD 45364, shown in Figure \ref{fig:HD45364.example_history}, the unstable simulations grow in libration amplitude and eventually escape from the resonance.
The unstable simulations for HD 33844, shown in Figure \ref{fig:HD33844.example_history}, exhibit more complicated behavior. 
Upon reaching the unstable ACR equilibrium, they initially grow in libration amplitude. 
Three of the simulations  go on to escape from the resonance, just as in the unstable HD 45364 simulations.
The other unstable simulations, however, reach a limit cycle. 
This behavior is consistent with the predictions of Figure \ref{fig:HD45364.acr_stability}, where many of the simulations lie in the pink shaded region of parameter space, leading to an unstable 2:1 MMR  but with ${\cal D}<{\cal D}_\text{bif}$.

    For HD 33844, Figure \ref{fig:HD45364.acr_stability} shows that there is a significant region of parameter space which is both consistent with our ACR model fit and for which the 2:1 ACR configuration is unstable but the 5:3 ACR configuration remains stable. HD 33844 b and c could, therefore, have readily evaded permanent capture in the 2:1 MMR and continued migrating toward the 5:3 MMR, where they were ultimately permanently captured.
    There is also a region of parameter space for HD 45364 that is consistent with our ACR model fit and for which the 2:1 ACR configuration is unstable but the 3:2 ACR configuration remains stable.
    HD 45364 b and c could, therefore, also have escaped the 2:1 MMR and continued migrating toward the 3:2 MMR.
    However, in order to migrate into the 3:2 MMR, the system would also have traversed the second-order 5:3 MMR. 
    There is a narrow region of parameter space in Figure \ref{fig:HD45364.acr_stability} that lead to unstable ACR equilibria in both the 2:1 and 5:3 MMRs but a stable 3:2 ACR equilibrium. 
    However, such parameters are not generic and a scenario in which the planet pair escaped both the 2:1 and 5:3 MMRs through instability but reached a stable equilibrium in the 3:2 MMR appears to suffer from a fine-tuning problem.
Alternatively, HD 45364 b and c could have reached the 3:2 MMR if their migration rate was sufficiently fast to avoid capture into the 5:3 MMR. 
This would require a migration rate sufficiently fast to avoid the 5:3 MMR but not too fast so as to avoid the 3:2 MMR. 
This is possible provided $\fracbrac{m_1+m_2}{M_*}^{-2} \lesssim \tau_\alpha / P \lesssim \fracbrac{m_1+m_2}{M_*}^{-4/3}$ \citep{Xu2017}.

How do our inferences about the migration histories of HD 45364 and HD 33844 compare to theoretical predictions?
For planets experiencing type I migration, the relative strength of eccentricity and migration rates, $\tau_\alpha/\tau_e$, are expected to be $\gtrsim 100$ \citep[e.g.,][]{Baruteau2014}, significantly larger than the values we infer in Figures \ref{fig:HD45364.migration_summary} and \ref{fig:HD45364.acr_stability}.
This is especially true for HD 45364, where our constraints are the most precise. 
We note that \citet{Delisle2015} examined HD 45364 and found that $\tau_{e,1}/\tau_{e,2}\gtrsim 6$ and 
$\tau_e/\tau_\alpha \sim 10$ based on the reported orbital solution of \citet{Correia2009}; in agreement with our results.  
However, the planets in both systems are sufficiently massive that they are expected to have open gaps in their protoplanetary disk, so the predictions of type I migration are probably not applicable \citep[e.g.,][]{Kley2012}. 

Hydrodynamical simulations of the migration of massive, resonant planet pairs in a gas disk have been studied by a number of authors \citep{Bryden2000,Kley2000,Snellgrove2001,Papaloizou2003,Kley2004,Sandor2007,Crida2008,Rein2010,Andre2016}.
\citet{Kley2004} infer an effective $\tau_\alpha/\tau_e$ of approximately $\sim1\text{--}10$ from their simulations of two giant planets in a disk by comparing the results of their hydrodynamic simulations with simple damped $N$-body simulations.
This result is in good agreement with the values inferred from our ACR fits for both systems.
\citet{Crida2008} and \citet{Morbidelli2007} note the importance of an inner disk to provide eccentricity damping for the innermost planet.  
Figure \ref{fig:HD45364.acr_stability} shows that both systems would have required the eccentricity damping experienced by the inner planet to be comparable in magnitude to the damping experienced by the outer planet if they were to have escaped the 2:1 MMR due to instability and yet reached a stable equilibrium in their present-day resonances.

While we have sketched plausible migration histories for the HD 45364 and HD 33844 systems that can explain how they might evade capture in a 2:1 MMR and reach their current configurations, we cannot rule out other scenarios. 
For example, the planets may have had significantly different masses at the time they encountered the 2:1 MMR and only have grown to their current masses after becoming trapped in the resonances they currently occupy.  
{Another possibility is that the disk conditions, and thus migration and eccentricity-damping rates, varied significantly between planets' encounters with the 2:1 MMR and their subsequent encounters with more closely spaced resonances.}
Additionally, our treatment of eccentricity damping and migration is likely oversimplified. 
In Equation \eqref{eq:capture:dissipation_forces}, we parameterized the dependence of planets' semi-major axis evolution under disk forces by including the parameter ``$p$," which describes the degree to which eccentricity-damping forces conserve angular momentum ($p=1$ corresponds to eccentricity damping at constant angular momentum).
As \citet{Xu2018} note, for type I migration, this treatment is only valid when eccentricities are less than the local disk aspect ratio, $e \ll H/r$, where $H$ is the disk scale height and $r$ the orbital radius. 
\citet{Xu2018} find that using a more realistic treatment of migration and eccentricity-damping forces in the type I regime generally leads to larger equilibrium eccentricities and more stable capture when compared to the approximate treatment we employ in Equation \eqref{eq:capture:dissipation_forces}. 
For giant planets that open a common gap (the regime likely occupied by  the planets we have considered here), the dependence of migration and eccentricity-damping forces on planet eccentricities is less well studied, though hydrodynamical simulations by \citet{Crida2008} suggest a potentially complicated dependence.
Ultimately, a comparative study of a larger sample of resonant systems both in and interior to the 2:1 MMR, in conjunction with theoretical investigations of planet--disk interactions in systems of resonant gap-opening planet pairs, could shed further light on why some systems become captured in the 2:1 MMR while others manage be captured into closer resonances.

%%%%%%%%%%%%%%%%%%%%%%%%%%%%%%%%%%%%%%%%%%%%
\subsection{Improving Constraints with Follow-up Observations}
\label{sec:discussion:followup}
%%%%%%%%%%%%%%%%%%%%%%%%%%%%%%%%%%%%%%%%%%%%

\begin{figure*}
    \centering
    \includegraphics[width=0.45\columnwidth]{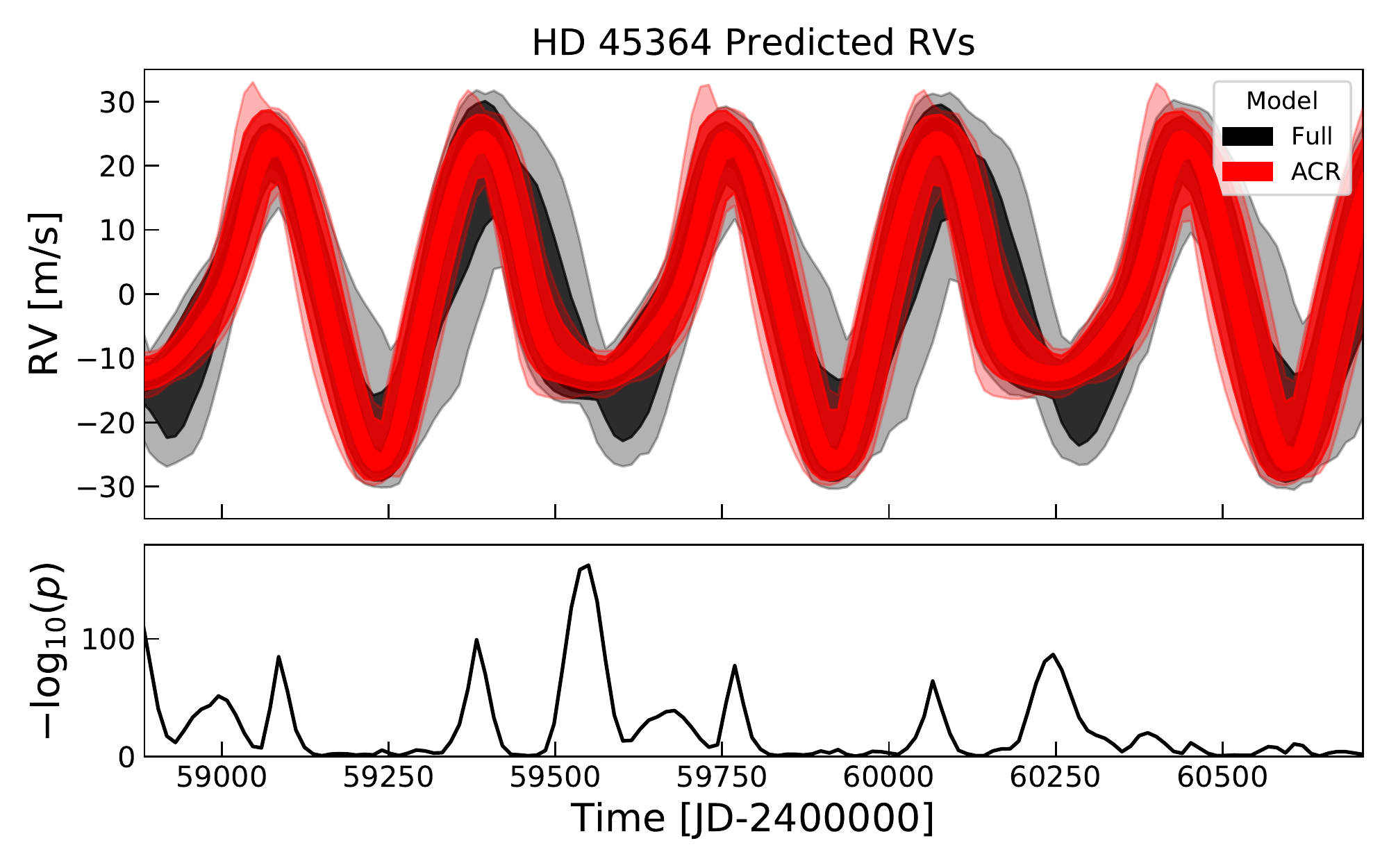}
    \includegraphics[width=0.45\columnwidth]{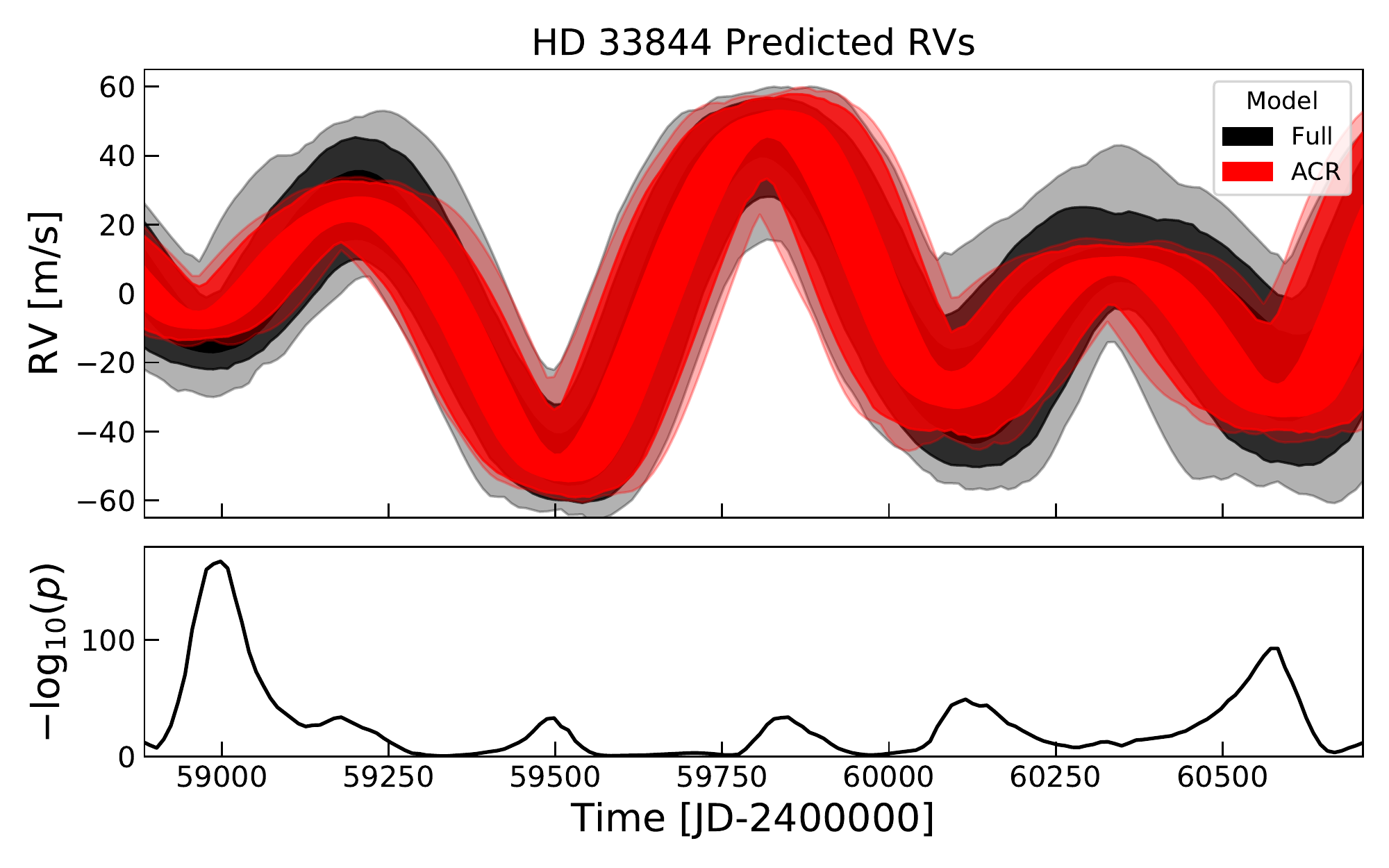}
    \caption{
    Predicted RV signals of the HD 45364  and HD 33844 system over the next 5 yr. 
    Top panels show RV signals for the unrestricted and ACR models.
    In the bottom panel, the times where the posterior predictive distributions differ between the unrestricted and ACR model most significantly are identified by computing $p$-values using a two-sample KS test for the two distributions.
    The predicted RVs of the ACR model have been computed via $N$-body integration as the double-Keplerian approximation does not remain valid over the timescales plotted (see Appendix \ref{app:validity}).
    }
    \label{fig:predicted.HD45364}
\end{figure*}

In Section \ref{sec:rv:data}, we showed that the RV signals of both HD 45364 and HD 33844 were consistent with the hypothesis that the systems reside in ACR configurations. 
In each case, our nested sampling simulations yielded model evidences that prefer ACR configurations for both systems. 
However, in both cases, the preference for the ACR models is somewhat marginal.
Here we briefly explore the prospects for follow-up observations to further bolster the evidence for ACR configurations in these systems or detect deviations from the predictions of the ACR models.
In Figure \ref{fig:predicted.HD45364}, we show the predicted RV signals of HD 45364 and HD 33844 from 2020 April 15 to 2025 April 15.
The figures compare the predicted RV signals projected using both the unrestricted and ACR models.
In order to identify the most opportune times for follow-up observations, we seek to identify times where the models' posterior predictive distributions of RV values are most distinct.
This is achieved by comparing the distribution of the predicted RV of a system at a given time using a two-sample Kolmogorov--Smirnov (KS) test \citep{kolmogorov:1933,smirnov1948}.
In the bottom panels of Figure \ref{fig:predicted.HD45364}  we plot the negative logarithm of the $p$ values returned by these KS tests as a function of time. 
Small $p$  values indicate that the distributions predicted by the two are highly unlikely to be derived from the same underlying distribution and provide a heuristic measure of the future epochs at which additional observations could most effectively distinguish between the two models.

\section{Summary and Conclusions}
\label{sec:conclusions}
In this paper we developed an RV-fitting method that allows us to fit the RVs of a pair of planets under the assumption that the pair resides in an ACR configuration.
This approach to RV fitting has a number of benefits. 
First, it allows one to test the hypothesis
that the present-day dynamical configuration arose from resonant capture during smooth migration by comparing fits with the ACR model to conventional RV fits. 
Second, it allows for the efficient identification of stable orbital configurations in strongly interacting systems.
Finally, for systems consistent with an ACR configuration, the eccentricities of the planets can used to infer constraints on the system's migration history. 

We applied this model to two systems hosting pairs of resonant giant planets, HD 45364 and HD 33844. 
We summarize our key conclusions:
\begin{enumerate}
    \item ACR configurations were preferred over an unrestricted $N$-body model for both HD 45364 and HD 33844 systems. 
    This lends support to the hypothesis that the resonances in these systems 
    were established via smooth migration. 
    \item Based on the present-day dynamical configurations of these systems, we used our ACR fits to measure the relative strengths of migration and eccentricity damping at the time these systems formed. In both cases, we found that the data are best explained by an eccentricity-damping timescale approximately $\sim10$ times shorter than the migration timescale.
    \item We showed that both systems could have avoided permanent capture in the 2:1 MMR during migration due to an instability mechanism while reaching stable equilibria in their present-day resonances. This can explain how these planets reached their present-day resonances without the need to invoke rapid migration in order to explain how they avoided permanent capture during earlier encounters with the 2:1 MMR.
\end{enumerate}
    We envision the application of the methods developed in this paper to the wider sample of resonant and near-resonant RV systems, allowing for a comprehensive comparative study, as a promising future direction for better understanding the role of migration in shaping the architectures of exoplanetary systems. 
    We provide the \texttt{Python} code for computing and fitting our ACR model online at \github.

\software{\texttt{exoplanet} \citep{exoplanet}, \texttt{dynesty} \citep{Speagle2020}, \texttt{radvel} \citep{radvel},  \texttt{REBOUND} \citep{rebound}, \texttt{REBOUNDx} \citep{reboundx}, \texttt{Theano}   \citep{theano}}
\acknowledgments
Acknowledgments.
We thank Paul Duffell, Matt Holman, and Josh Speagle for helpful discussion.
S.H. gratefully acknowledges the CfA Fellowship.
{We are grateful to the anonymous referee whose feedback has improved the quality of this manuscript.} The computations in this paper were run on the Odyssey cluster supported by the FAS Science Division Research Computing Group at Harvard University.
\appendix
%%%%%%%%%%%%%%%%%%%%%%%%%%%%%%%%%%%%%%%%%%%%%%%%%%%%%%%%%%%%%
%%%%%%%%%%%%%%            Appendix         %%%%%%%%%%%%%%%%%
%%%%%%%%%%%%%%%%%%%%%%%%%%%%%%%%%%%%%%%%%%%%%%%%%%%%%%%%%%%%%
\section{Equations of Motion}
\label{app:Equations}
\subsection{Construction of the Hamiltonian}
\label{app:Equations:hamiltonian}
In this appendix  we derive the equations of motion governing the dynamics of a $j$:$j-k$ resonance between a pair of coplanar planets of mass $m_1$ and $m_2$ orbiting a star of mass $m_0$. 
We begin by considering the conservative evolution of the system in Jacobi coordinates using modified canonical Delauney variables \citep[e.g.,][]{MorbidelliBook2002}. 
The canonical momenta are given by
\begin{align*}
\Lambda_i = {\tilde{m}_i}\sqrt{G\tilde{M}_i a_i}\\
\Gamma_i = {\tilde{m}_i}\sqrt{G\tilde{M}_i a_i}(1-\sqrt{1-e_i^2})
\end{align*}
where $G$ is the gravitational constant, $\tilde{m}_i=\frac{\eta_{i-1}}{\eta_i}m_i$, 
$\tilde{M}_i = \frac{\eta_{i}}{\eta_{i-1}}m_0$,
and $\eta_i=\sum_{j=0}^{i}m_i$.
The angle variables conjugate to $\Lambda_i$ and $\Gamma_i$ are are, respectively, the planets' mean longitudes,  $\lambda_i$, 
and $\gamma_i=-\varpi_i$ where $\varpi_i$ are the planets' longitudes of periapse. 
The Hamiltonian is given by 
\begin{eqnarray}
H(\lambda_i,\gamma_i;\Lambda_i,\Gamma_i)=-\sum_{i=1}^2\frac{G^2{\tilde M}_i^2{\tilde m}_i^3}{2\Lambda_i^2}-Gm_1m_2\left(\frac{1}{|r_1-r_2|}-\frac{r_1\cdot r_2}{|r_2|^3} \right) 
\end{eqnarray}
where the position vectors $r_i$ are functions of the canonical variables, and we omit terms that are second order and higher in the planet--star mass ratios. 

To derive equations of motion governing the dynamics of a $j$:$j-k$ resonance, we follow \citet{Michtchenko2006} and transform to new canonical angle variables,
\begin{align}
\begin{pmatrix}\sigma_1\\ \sigma_2 \\ Q \\ l \end{pmatrix}
=
\begin{pmatrix} 
-s & 1+s & 1 & 0 \\
-s & 1+s & 0 & 1 \\
-1/k & 1/k & 0 & 0 \\
\frac{1}{2} & \frac{1}{2} & 0 & 0 
\end{pmatrix}
\cdot
\begin{pmatrix}\lambda_1 \\ \lambda_2 \\ \gamma_1 \\ \gamma_2 \end{pmatrix}
\label{eq:equations_of_motion:angles}
\end{align}
where $s = (j-k)/{k}$, along with new conjugate momenta defined implicitly in terms of the old momentum variables as
\begin{eqnarray}
\Gamma_i &=& I_i \nonumber\\
\Lambda_1 &=& \frac{L}{2} - P/k - s (I_1 + I_2) \nonumber \\
\Lambda_2 &=& \frac{L}{2} + P/k + (1+s) (I_1 + I_2)~.\label{eq:equations_of_motion:momenta}
\end{eqnarray}
The full Hamiltonian is independent of the new angle $l=(\lambda_1+\lambda_2)/2$ and its corresponding canonical momentum,
$$
L = \Lambda_1 - \Gamma_1 + \Lambda_2 -\Gamma_2~,
$$
is the total angular momentum, which is conserved due to the rotational symmetry of the system. 

To derive equations of motion governing the resonant dynamics, we perform a near-identity canonical transformation $(\sigma_i,I_i,Q,P)\rightarrow(\bar\sigma_i,\bar I_i,\bar Q,\bar P)$
such that, to leading order in the planet--star mass ratio, 
the new Hamiltonian is given by
\begin{equation}
    {\bar H}(\bar\sigma_i, \bar I_i; L, \bar P) = \frac{1}{2\pi}\int_{0}^{2\pi} H dQ~,
    \label{eq:equations_of_motion:Hbar}
\end{equation}
i.e., the new Hamiltonian is the old Hamiltonian averaged over the ``fast'' variable $Q$.
We construct this transformation explicitly below in Appendix \ref{app:Equations:transformation} so that we may transform the osculating elements of a pair of planets to the variables of our Hamiltonian model and vice versa. 
%
%$\bar x = \exp[{\cal L}_{\chi_1}()]x$

The averaging procedure yields
$$
{\bar P}/k= \frac{\bar\Lambda_2-\bar\Lambda_1}{2}-\left(s+\frac{1}{2}\right)(\bar\Gamma_1+\bar\Gamma_2)
$$
as a constant of motion.
 In order to gain a better physical intuition for the meaning of the conserved quantity, $\bar P$, it is instructive to express its value in terms of the value of the planets' eccentricities and distance to nominal resonance, $\frac{j-k}{j}\frac{P_2}{P_1}-1\equiv\Delta$.
This can be accomplished as follows:
first, define the reference semi-major axes $a_{2,0}$ 
and $a_{1,0} = \fracbrac{\tilde{M}_1}{\tilde{M}_2}^{1/3}\fracbrac{j-k}{j}^{2/3}a_{2,0}$ corresponding to the nominal location of resonance, such that $ L= \Lambda_{1,0} +\Lambda_{2,0}$ where
$\Lambda_{i,0}=\tilde{m_i}\sqrt{G\tilde{M}_i a_{i,0}}$ for $i=1,2$.
Conservation of $\bar P$ and $L$ implies that $\bar \Lambda_{2}-\Lambda_{2,0}=-\frac{s+1}{s}(\bar \Lambda_{1}-\Lambda_{1,0})$.
Defining $\delta\Lambda_{i} = \Lambda_{i}-\Lambda_{i,0}$,
we can write
\begin{eqnarray*}
 \Delta \approx3\left(\frac{\delta \Lambda_2 }{\Lambda_{2,0}}-\frac{\delta \Lambda_1 }{\Lambda_{1,0}}\right)
    =3\delta \Lambda_2 \left(\frac{(s+1)\Lambda_{1,0} + s\Lambda_{2,0}}{(s+1)\Lambda_{2,0}\Lambda_{1,0}}\right)~.
\end{eqnarray*}
We define the `normalized AMD'  as
\begin{eqnarray}
    {\cal D}&=&\frac{\frac{1}{2}(\Lambda_{2,0}-\Lambda_{1,0})-\bar P/k}{(s+1/2)(\tilde{m}_1+\tilde{m}_2)\sqrt{Gm_0a_{2,0}}}~.
    \label{eq:normalized_amd_defn}
\end{eqnarray}
Rewriting Equation \eqref{eq:normalized_amd_defn} in terms of orbital elements gives 
\begin{equation*}
    {\cal D}\approx
    \beta_{1}\sqrt{\alpha_0}\left(1-\sqrt{1-e_1^2}\right)
    +
    \beta_{2}\left(1-\sqrt{1-e_2^2}\right)
    -\frac{k\beta_2\beta_1 \sqrt{\alpha_0} \Delta }{3 \left(\beta_1\sqrt{\alpha_0}  (s+1)+\beta_2 s\right)}
\end{equation*}
where $\beta_i=\frac{\tilde{m}_i}{\tilde{m}_1+\tilde{m}_2}\sqrt{\frac{{\tilde M}_i}{m_0}}$ and $\alpha_0=a_{1,0}/a_{2,0}$.
For small eccentricities and $\Delta\approx 0 $, we have that ${\cal D}\approx\beta_1\sqrt{\alpha_i}e_1^2/2 + \beta_2e_2^2/2$.

In order to implement the equations of motion given in Equation \eqref{eq:capture:equations_of_motion} governing resonant dynamics,  we evaluate the integral in Equation \eqref{eq:equations_of_motion:Hbar} and its derivative with respect to canonical variables by numerical quadrature using a Gauss-Legendre quadrature rule.  
We use the \texttt{exoplanet} \citep{exoplanet} package's Kepler solver algorithm to compute the planets' $\vec{r}_i$ as functions of canonical variables to evaluate the integrand of Equation \eqref{eq:equations_of_motion:Hbar}. Derivatives of $\bar{H}$ are then computed using the \texttt{Theano} \citep{theano} package's automatic differentiation capabilities. 
%
%%%%%%%%%%%%%%%%%%%%%%%%%%%%%%%%%%
%%%%%%% osculating elements %%%%%%
%%%%%%%%%%%%%%%%%%%%%%%%%%%%%%%%%%
\subsection{Transformation from mean to osculating elements}
\label{app:Equations:transformation}
The orbital elements appearing in the canonical variables of the equations of motion governing the planets' resonant dynamics represent `mean' elements.
These mean elements differ from the planets' osculating elements, which exhibit additional high-frequency variations on the timescale of the planets' synodic period.
This is because the Hamiltonian system governing the resonant dynamics is derived from a phase-space reduction of the full three-body problem via near-identity canonical transformation that eliminates the degree of freedom associated with the canonical coordinate $Q = (\lambda_2-\lambda_1)/k$.
The ACR equilibrium configurations of the resonant 
Hamiltonian correspond to periodic orbits of the full three-body problem that are $2\pi$-periodic in $Q$.
In order to properly initialize $N$-body simulations with planets residing in an ACR configuration, the periodic variations of the osculating orbital elements must be accounted for;
ignoring these corrections can result in large libration amplitudes when running $N$-body simulations.
We do this by constructing the canonical transformation from the variables of our resonance model to the canonical coordinates of the unaveraged three-body problem.

In order to construct our canonical transformation, we begin by writing the full Hamiltonian of the planar three-body problem  as 
\begin{eqnarray}
    H(\sigma_i,I_i,Q,P; L) &=& H_0(I_i,P; L) + \epsilon \bar{H}_1(\sigma_i,I_i,P; L) + \epsilon H_{1,\text{osc.}}(\sigma_i,I_i,Q,P; L)
\end{eqnarray}
where
\begin{eqnarray}
H_0&=&-\sum_{i=1}^2\frac{G^2{\tilde M}_i^2{\tilde m}_i^3}{2\Lambda_i^2}\nonumber\\
   \epsilon \bar{H}_1 &=& \frac{Gm_1m_2}{2\pi}\int_{0}^{2\pi}\left(\frac{1}{|r_1-r_2|}-\frac{r_1 \cdot r_2}{r_2^3}\right)dQ\nonumber\\
   \epsilon  H_{1,\text{osc.}}&=&Gm_1m_2\left(\frac{1}{|r_1-r_2|}-\frac{r_1 \cdot r_2}{r_2^3}\right) - \bar{H}_1
\end{eqnarray}
and $\epsilon$ serves as a bookkeeping parameter that we will set to unity after deriving the transformation to first order in $\epsilon$.
We focus on constructing the transformation for phase-space points in the vicinity of an ACR. 
We adopt the vector notation ${\bf z}=(\sigma_1,\sigma_2,I_1,I_2)$ and define an equilibrium point $\bar{\bf z}_\text{eq}(P_0)$ as a point that satisfies
\begin{equation*}
    \nabla_{\bf z}{\left[H_0(\bar{\bf z}_\text{eq},P_0;L) + \epsilon \bar{H}_1(\bar{\bf z}_\text{eq},P_0;L)\right]} = 0
\end{equation*}
We next transform to canonical variables 
$\delta {\bf z} = {\bf z} - \bar{\bf z}_\text{eq}$ and $\delta P = P-P_0$ centered on the equilibrium configuration
and approximate the Hamiltonian by expanding in $\delta {\bf z}$ as
\begin{eqnarray}
H &=&\frac{1}{2}\delta {\bf z}^{T} \cdot \nabla^2_{\bf z}\left[H_0 + \epsilon \bar{H}_1\right]\cdot \delta {\bf z}  +\left(\omega_\text{syn}+\delta{\bf z}\cdot\nabla_{\bf z}\omega_\text{syn}\right)\delta P/k 
% + \epsilon \bar{H}_1(\bar{\bf z}_\text{eq},P)
+ \epsilon\left[H_{1,\text{osc.}} +\delta{\bf z}^{T}\cdot\nabla_{\bf z} H_{1,\text{osc.}} \right] \label{eq:Hexpand}
\end{eqnarray}
where $\omega_\text{syn}\equiv k\pd{H_0}{P}$ is the synodic frequency $n_2-n_1$, and we have dropped constant terms from Equation \eqref{eq:Hexpand}. 
We next perform a symplectic linear transformation, 
\begin{equation}
    {\bf w} = S^{-1}\cdot \delta{\bf z}\label{eq:z_to_w}
\end{equation}
to new canonical variables ${\bf w} = (w_1,w_2,\tilde{w}_1,\tilde{w}_2)$ where $w_i$ are the new coordinates and $\tilde{w}_i$ are the conjugate momenta. 
The matrix $S$ satisfies
\begin{eqnarray}
    S^{-1} \cdot \left(\Omega\cdot \nabla^2_{\bf z}\left[H_0 + \epsilon \bar{H}_1\right]\right)\cdot S = D\nonumber\\
    S^{T}\cdot \Omega \cdot S = \Omega
\end{eqnarray}
where $D$ is a diagonal matrix of the form $D = \text{diag}(i\omega_1,i\omega_2,-i\omega_1,-i\omega_2)$.
We choose $S$ so that the components of $\bf w$ satisfy  $\tilde{w}_i = -i w_i^*$.
After the canonical change of variables, Hamiltonian \eqref{eq:Hexpand} transforms to 
\begin{equation}
H({\bf w},\delta P, Q) = i\sum_{i=1}^{2}\omega_i w_i\tilde{w}_i  + \left(\omega_\text{syn}+ {\bf w}\cdot\nabla_{\bf w}\omega_\text{syn}\right)\delta P/k +  \epsilon\left[H_{1,\text{osc.}}(\bar{\bf z}_\text{eq},Q) +{\bf w} \cdot\nabla_{\bf w} H_{1,\text{osc.}}(Q) \right]
\label{eq:ham_w_no_av}
\end{equation}
where $\nabla_{\bf w} = S^T\cdot \nabla_{\bf z}$.
This transformation puts the Hamiltonian in the form of a perturbed pair of harmonic oscillators with frequencies $\omega_i$ and allows us to apply canonical perturbation theory to construct the transformation from osculating to mean variables below.
To construct the transformation from $\delta {\bf z}$ to $\bf w$, we evaluate the matrix $\Omega\cdot\nabla^2_{\bf z}\left[H_0 + \epsilon \bar{H}_1\right]$ by means of automatic differentiation and then  numerically determine its eigenvectors in order to calculate the matrices $S$ and $D$.

We now construct a canonical transformation $({\bf w},\delta P, Q)\rightarrow(\bar{\bf w},\delta \bar P,\bar Q)$ by means of a canonical Lie transformation \citep[e.g.,][]{MorbidelliBook2002} 
in order to eliminate the dependence of the Hamiltonian on $Q$ to first order in $\epsilon$. The transformation is generated by the function $\chi_1(\bar{\bf w},\delta \bar P,\bar Q)$ so that $({\bf w},\delta P, Q) = \exp{[\epsilon\mathcal{L}_{\chi_1}]}(\bar{\bf w},\delta \bar P,\bar Q)$,
where $\mathcal{L}_{\chi_1} = \left\{\cdot,\chi_1\right\}$ is the Lie derivative with respect to $\chi_1$, i.e., 
the canonical Poisson bracket of a given function of phase-space coordinates and the function $\chi_1$.
We will solve for $\chi_1$ so that Hamiltonian \eqref{eq:ham_w_no_av} is transformed to  
\begin{eqnarray}
  \bar{H}(\bar{\bf w},\delta \bar P,\bar Q) &\equiv&
  \exp{[\epsilon\mathcal{L}_{\chi_1}]}H(\bar{\bf w},\delta \bar P,\bar Q)=
  i\sum_{i=1}^{2}\omega_i \bar{w}_i\bar{\tilde{w}}_i +\left(\omega_\text{syn}+ \bar{\bf w}\cdot\nabla_{\bf w}\omega_\text{syn}\right)\delta \bar P/k + {\cal O}(\epsilon^2)
  \label{eq:ham_transformed}~.
\end{eqnarray}
Expanding Equation \eqref{eq:ham_transformed} 
and collecting terms that are first order in $\epsilon$, we obtain the following equation for $\chi_1$:\footnote{As $\bar P$ is a conserved quantity of the resonant Hamiltonian, $P-P_0$ will be ${\cal O}(\epsilon)$.
The term $(\delta P/k)\left\{\bar{\bf w}\cdot\nabla_{\bf w}\omega_\text{syn} ,\chi_1\right\}$, is therefore omitted from Equation \eqref{eq:chi_equation} as it is ${\cal O}(\epsilon^2)$}
\begin{eqnarray}
i\sum_{i=1}^{2}\omega_i\left(\bar{\tilde{w}}_i\pd{\chi_1}{\bar{\tilde{ w}}_i}-\bar{w}_i\pd{\chi_1}{\bar{w}_i}\right) + \frac{1}{k}\left(\omega_\text{syn} + \bar{\bf w}\cdot\nabla_{\bf w}\omega_\text{syn} \right)\pd{\chi_1}{\bar Q} 
+ H_{1,\text{osc.}}+\bar {\bf w} \cdot\nabla_{\bar{\bf w}} H_{1,\text{osc.}}=0~.
\label{eq:chi_equation}
\end{eqnarray}
Inserting the ansatz solution 
\begin{equation}
    \chi_1(\bar{Q},\bar{w}_i,\bar{\tilde{w}}_i) = G(\bar Q) +
    \sum_i \bar{w}_i F_i(\bar{Q}) +
    \bar{\tilde{w}}_i\tilde{F}_i(\bar{Q})
    \label{eq:chi_ansatz}
\end{equation}
for $\chi_1$ into Equation \eqref{eq:chi_equation}, we have
\begin{eqnarray}
\frac{\omega_\text{syn}}{k}\dd{G(Q)}{Q}  &=& H_{1,\text{osc.}}(\bar{\bf z}_\text{eq},Q)\label{eq:chi_ode1}\\
\frac{\omega_\text{syn}}{k}\dd{F_i(Q)}{Q}  &=&\pd{H_{1,\text{osc.}}(Q)}{w_i}-\frac{1}{k}\pd{\omega_\text{syn}}{w_i}\dd{G(Q)}{Q}+ i\omega_i F_i(Q)
\label{eq:chi_odes}
\end{eqnarray}
and $\tilde F_i = -iF_i^*$.
To solve for $F_i$, we Fourier decompose $F_i$ and $H_{1,\text{osc.}}$ as
$F_i = \sum_{l}\hat F_{i,l}\exp[ilQ]$ and
\begin{eqnarray*}
\left[\pd{}{\bar{w}_i}-\pd{\ln\omega_\text{syn}}{\bar{w}_i}\right]H_{1,\text{osc.}}(\bar{\bf z}_\text{eq},Q)= \sum_{l}\hat X_{i,l}\exp[ilQ]~.
\end{eqnarray*}
We use a  fast Fourier transform algorithm to compute the amplitudes $\hat X_{i,l}$ and 
then compute solutions for $F_1(Q)$ and $F_2(Q)$ as Fourier sums with amplitudes
\begin{equation}
    {\hat F_{i,l}} = \frac{-ik}{l\omega_\text{syn}-k\omega_i}\hat{X}_{i,l}~.
\end{equation}

\begin{figure}
    \centering
    \includegraphics[width=0.6\textwidth]{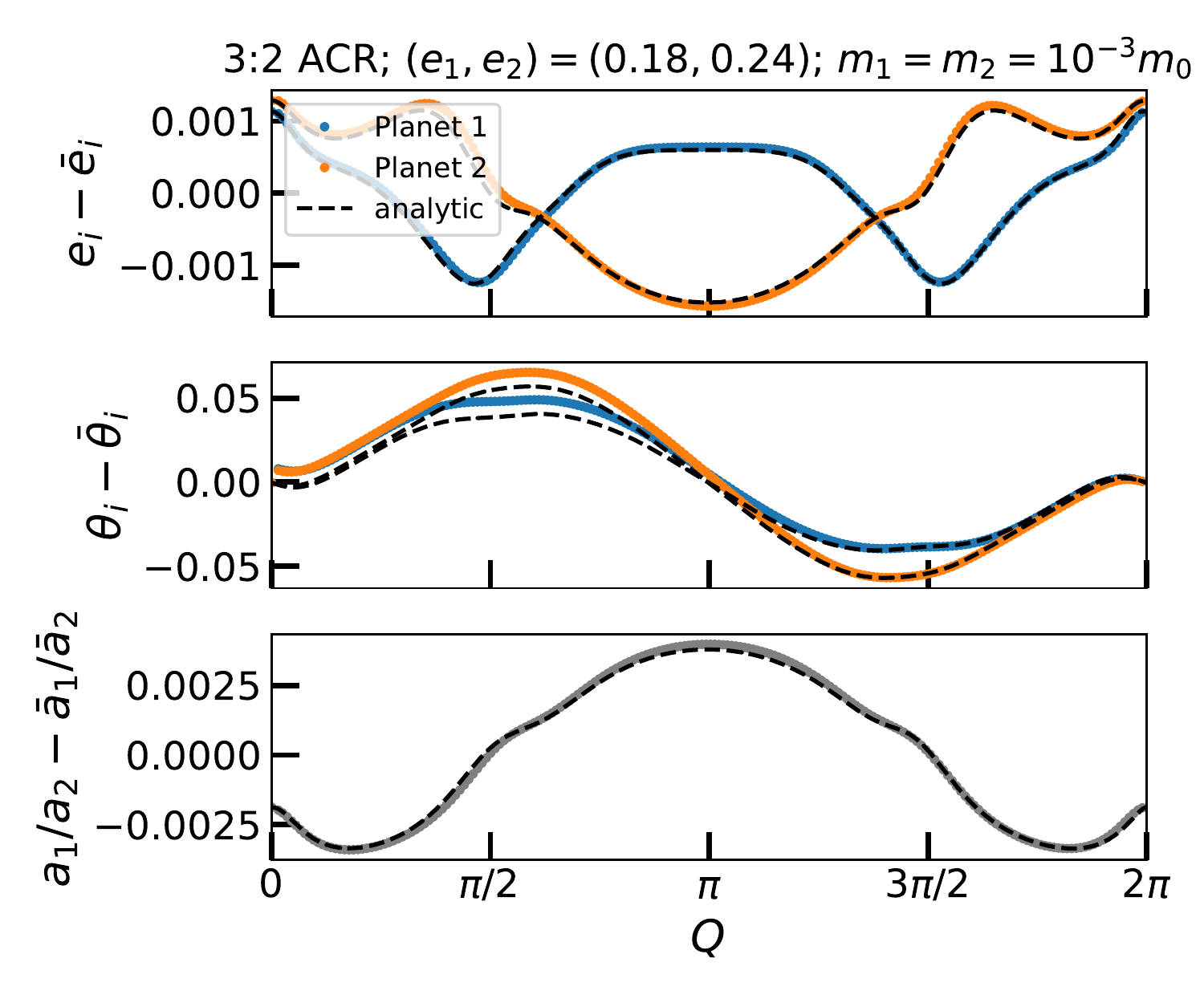}
    \caption{
    Difference between osculating orbital elements and mean orbital elements computed from the canonical variables of our resonance model Hamiltonian as a function of $Q$ for a pair of planets with $m_1=m_2 = 10^{-3}m_0$.
    Black dashed lines show analytic predictions computed using Equations \eqref{eq:osc_P_soln} and \eqref{eq:osc_z_soln} while results of an $N$-body integration are shown as colored points.
    The $N$-body integration was initialized by setting the planets' orbital elements to the analytically computed osculating elements for $Q=0$ and integrated until $Q=2\pi$.
    The particular ACR configuration shown corresponds to ${\cal D}=0.22$.
     Numerical routines for reproducing this figure
    as well as computing the transformation from mean to osculating variables for different resonances, planet masses, and normalized AMD values are available online at \github.
    }
    \label{fig:osc_correction_example}
\end{figure}

We convert the canonical variables of the ACR equilibrium solutions of the resonance Hamiltonian to the canonical variables of the unaveraged three-body problem which are then, in turn, used to compute the osculating orbital elements of planets in the ACR configuration.
To first order in $\epsilon$, this is achieved by evaluating the function $\bar x \mapsto x= \bar x + \{\bar x,\chi_1\}$ at the phase-space point $({\bar{\bf w}},\delta \bar{P},\bar Q) =(0,0,Q)$
for any dynamical variable or function of dynamical variables, $x$, of interest.  
For a given ACR equilibrium solution of the resonance Hamiltonian, $\bar {\bf z}_\text{eq}(P_0)$,  Equations \eqref{eq:z_to_w}, \eqref{eq:chi_ansatz}, and \eqref{eq:chi_ode1}, give the solutions
\begin{eqnarray}
    P(Q)  &=& P_0+\left\{\delta\bar P,\chi_1 \right\}_{\bf{\bar w} = 0} = P_0 + \frac{k}{\omega_\text{syn}}H_{1,\text{osc.}}(\bar{\bf z}_\text{eq},Q)~\label{eq:osc_P_soln}\\
{\bf z}(Q) &=&  \bar{\bf z}_\text{eq} + \left\{\delta\bar{\bf   z},\chi_1\right\}
=  \bar{\bf z}_\text{eq} + S \cdot \Omega \cdot \begin{pmatrix}
F_1(Q)\\
F_2(Q)\\
-iF^*_1(Q)\\
-iF^*_2(Q)
\end{pmatrix}\label{eq:osc_z_soln}
\end{eqnarray}
    for osculating canonical variables, which can then be used to compute the planets' osculating orbital elements.
    Figure \ref{fig:osc_correction_example} shows an example of osculating elements computed as a function of $Q$ using the transformations given in Equations \eqref{eq:osc_P_soln} and \eqref{eq:osc_z_soln} for a pair of Jupiter-mass planets in a 3:2 ACR.
    Numerical routines for reproducing Figure \ref{fig:osc_correction_example}
    as well as computing the transformation from mean to osculating variables for different resonances, planet masses, and normalized AMD values are available online at \github.
    
\section{Validity of the Two-Keplerian Approximation}
\label{app:validity}
    While our unrestricted model utilizes $N$-body integration to synthesize RV signals, the ACR model relies on approximating the RV signal as the sum of two independent signals generated by two planets on strictly Keplerian orbits.
    Gravitational interactions between the planets cause their orbits to deviate from perfect Keplerian ellipses and modify the RV signal. 
    Here we briefly explore the consequences of these effects on the accuracy of our ACR model for the planets' RV signal.

    The most important consequence of resonant planets' gravitational interactions is that they induced periapse precession at a constant rate. (The periodic variations of the osculating elements described above in Appendix \ref{app:Equations:transformation} have minimal influence on the RV signal.)
    Due to this precession, the average value of resonant planets' period ratio deviates slightly from the nominal resonant value. 
    For example, if HD 45364 b and c are in a 3:2 ACR, they will have orbital periods such that  $3\dot{\lambda}_c - 2\dot{\lambda}_b$ is equal to the mean apsidal precession rate, $\dot{\varpi}_b=\dot{\varpi}_c$, rather than 0.
    This ensures that the average values of the resonant angles remain fixed at their equilibrium values.
    We have ignored this effect in our two-Keplerian approximation and instead set the planets' period ratios to their nominal resonant ratio. We also treat planets' longitudes of periapse as constant. 
    
\begin{figure}
        \centering
        \includegraphics[width=0.75\textwidth]{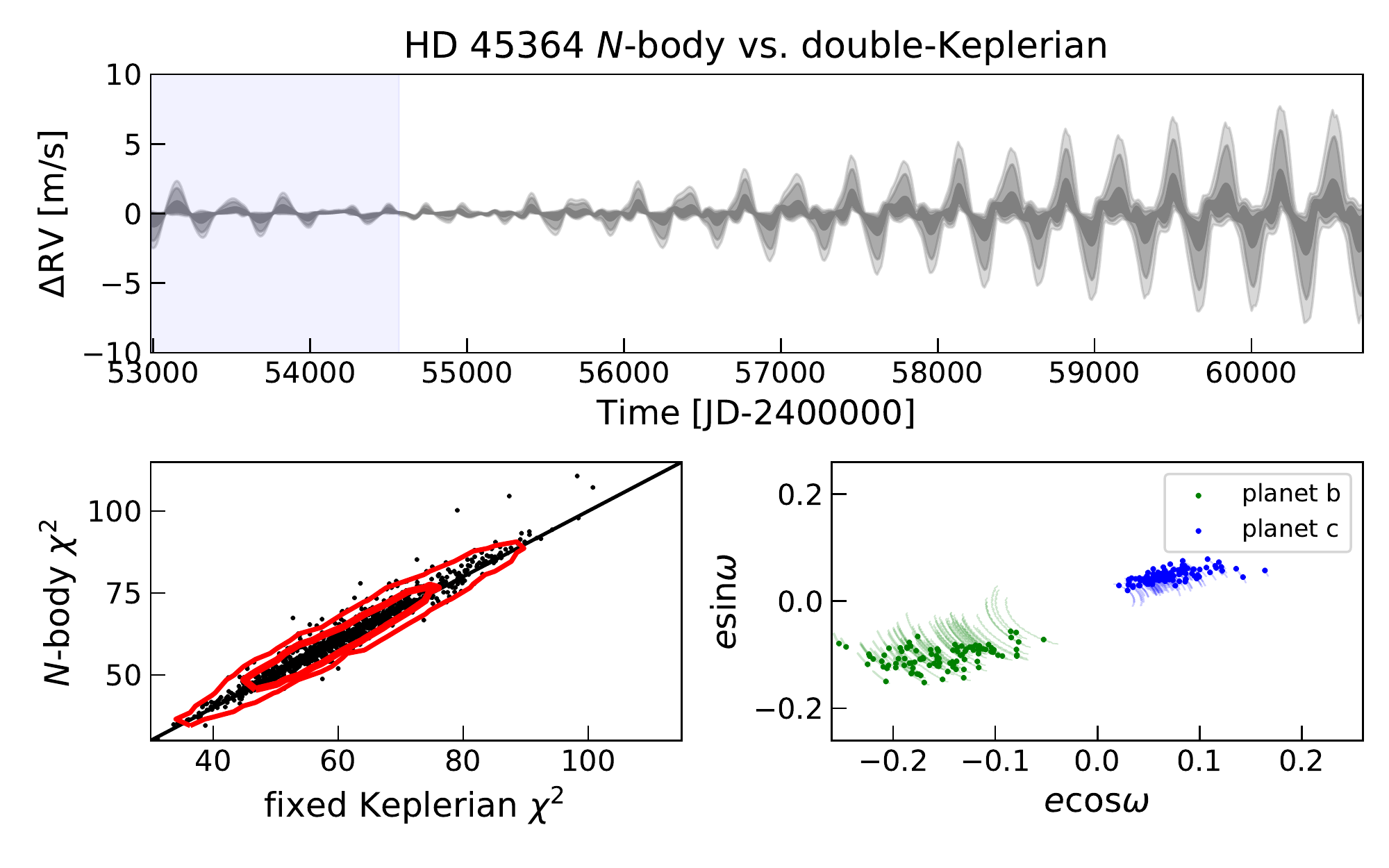}
        \includegraphics[width=0.75\textwidth]{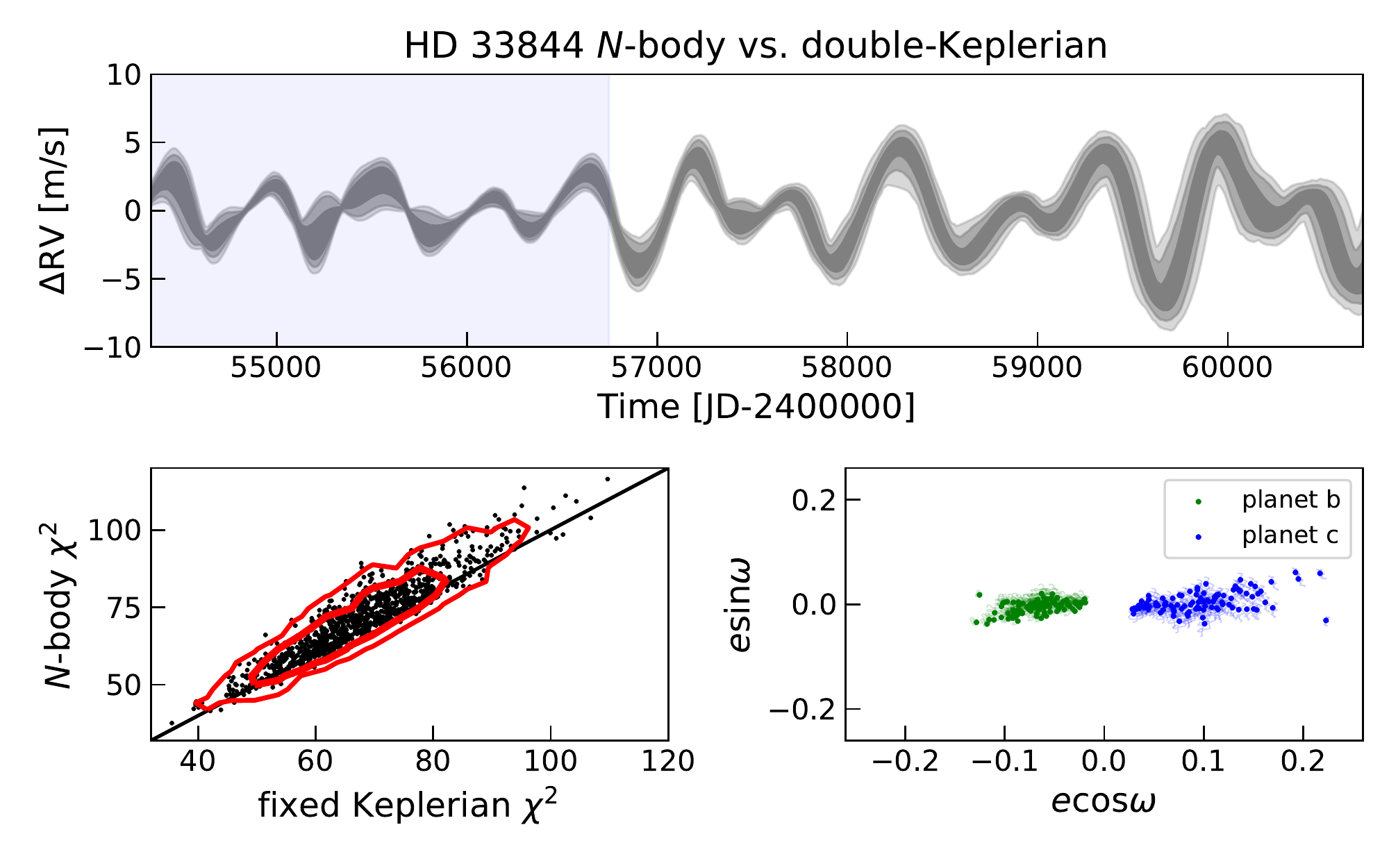}
        \caption{
        Comparison between RV signals of resonant configurations computed using $N$-body integrations versus a double-Keplerian approximation.
        For each system, the top panels show the velocity predicted via $N$-body minus the velocity predicted by the double-Keplerian model for 1000 random draws from the MCMC posterior fit.
        The shading indicates the regions containing 68\%, 95\%, and 99.7\% ($1\sigma$,$2\sigma$, and $3\sigma$) of the samples' velocity differences as a function of time. 
        The shaded blue regions indicate the time spanned by the observational data reported by \citet{Correia2009} for HD 45364 and \citet{Wittenmyer2016} for HD 33844.
        The bottom-left panels show scatter 
        plots of the $\chi^2$ value computed with $N$-body integrations versus the $\chi^2$ computed when the double-Keplerian model is used.
        The right panels illustrate the precession of the planets' eccentricity vectors.
        The fixed eccentricity and $\omega$ values assumed by the double-Keplerian model are plotted as points with a line showing the precession of the eccentricity vectors over the course of the plotted time frame.
        Parameters of the two-Keplerian ACR models are converted to  an $N$-body simulation assuming the planets are coplanar in an $i=90^\circ$ edge-on configuration.
        The planets' orbital elements are translated from the mean elements of the ACR model to osculating elements in the $N$-body simulations using the procedure described in Appendix \ref{app:Equations}.
        }
        \label{fig:nbody_vs_fixed}
    \end{figure}

    Figure \ref{fig:nbody_vs_fixed} plots the difference between 
    the RV signals of planets in an ACR configuration computed with the fixed Keplerian approximation versus the 
    RV signals of planets in ACR configurations computed by $N$-body integration.
    We utilize the transformations from mean to osculating elements described in Section \ref{app:Equations:transformation} in order to initialize the $N$-body simulations with planets in ACR configurations.
    Figure \ref{fig:nbody_vs_fixed} shows that, at the time of the originally reported RV observations, differences between $N$-body and double-Keplerian models are small and should not affect our conclusions, 
    given that the median RV measurement uncertainty, including the inferred jitter, is $\sim$2 m/s for HD 45364 and $\sim$8 m/s for HD 33844.

    % The two-Keplerian approximation slightly miscalculate the phases the line of apses will rotate with respect to an observer's of sight and the 

\bibliographystyle{yahapj}
\bibliography{main.bib}
\end{document}